\pdfoutput=1
\documentclass[12pt]{article}

\usepackage[left=0.8in,right=0.8in,top=1.0in,bottom=1.0in]{geometry}

\usepackage{hyperref}
\usepackage{color}
\usepackage{amsmath,amssymb}
\usepackage{bm}
\usepackage{graphicx, color}
\usepackage{wrapfig}
\usepackage{here}
\usepackage{cite}
\usepackage{ulem}
\usepackage{mathtools}
\usepackage{here}

\usepackage{multirow}

\newcommand{\CL}[2]{\mathcal{C}_{\underset{#2}{#1}}}

\newcommand{\brackets}[1]{\left( #1 \right)}
 
\newcommand{\cO}{{\mathcal O}}

\definecolor{green}{rgb}{0, 0.5, 0} 

\numberwithin{equation}{section}

\begin{document}

\title{
	\begin{flushright}
		\ \\*[-80pt]
		\begin{minipage}{0.22\linewidth}
			\normalsize
			EPHOU-22-008\\
			KYUSHU-HET-239\\*[50pt]
		\end{minipage}
	\end{flushright}
	{\Large \bf
		Lepton flavor violation, lepton $(g-2)_{\mu,\,e}$ 
		and electron EDM \\ in the modular symmetry
		\\*[20pt]}}

\author{ 
	\centerline{
		Tatsuo Kobayashi $^{1}$,   Hajime Otsuka $^{2}$,
		Morimitsu Tanimoto $^{3}$,  Kei Yamamoto  $^{4}$
	} 
	\\*[20pt]
	\centerline{
		\begin{minipage}{\linewidth}
			\begin{center}
				$^1${\it \normalsize
					Department of Physics, Hokkaido University, Sapporo 060-0810, Japan} \\*[5pt]
				$^2${\it \normalsize
					Department of Physics, Kyushu University, 744 Motooka, Nishi-ku, Fukuoka,
					819-0395, Japan}\\*[5pt]
				$^3${\it \normalsize
					Department of Physics, Niigata University, 
					Ikarashi 2-8050, Niigata 950-2181, Japan} \\*[5pt]
				$^4${\it \normalsize
					Department of Global Environment Studies,
					Hiroshima Institute of Technology, Hiroshima 731-5193, Japan} \\*[5pt]
			\end{center}
	\end{minipage}}
	\\*[80pt]}
\date{
	\centerline{\small \bf Abstract}
	\begin{minipage}{0.9\linewidth}
		\medskip 
		\medskip 
		\small
		\mbox{}
		We  study   
		the lepton flavor violation (LFV),  the leptonic magnetic moments
		$(g-2)_{\mu,\,e} $ and the electric dipole moment (EDM) of the electron in the Standard-Model Effective Field Theory
		with the $\Gamma_N$ modular flavor symmetry. 
		We employ the stringy Ansatz on coupling structure that 
		4-point couplings  of matter fields are written by 
		a product of 3-point couplings  of matter fields.  
		We  take the level 3 finite modular group, 
		$\Gamma_3$ for the flavor symmetry, and 
		discuss the dipole  operators  at nearby fixed point $\tau=i$,
		where observed  lepton masses and mixing angles are well reproduced.
		Suppose the anomaly of the anomalous magnetic moment of the muon
		to be {evidence} of the new physics (NP), we have related it with  $(g-2)_{e}$,  LFV decays, and the electron EDM. 
		It is found that the NP contribution to $(g-2)_{e}$ {is} proportional to the lepton {masses squared} likewise  the naive scaling.
		We also discuss the correlations among
		the LFV processes $\mu\to e\gamma$, $\tau\to \mu\gamma$ 
		and $\tau\to e\gamma$, which are testable in the future.
	The electron EDM requires the tiny imaginary part of
		the relevant  Wilson coefficient in the basis of real positive charged lepton masses, which
		is related to the $\mu\to e\gamma$
		transition in our framework.			
	\end{minipage}
}
\begin{titlepage}
	\maketitle
	\thispagestyle{empty}
\end{titlepage}


\section{Introduction}
The electric and  magnetic dipole moments  of 
the electron and the muon are  low-energy probes 
of {New Physics (NP)} beyond the Standard Model (SM). The recent experimental measurement of the anomalous magnetic moment of the muon, $a_\mu$
has indicated  the discrepancy with the SM prediction 
in Ref.~\cite{Muong-2:2021ojo,Muong-2:2006rrc,Aoyama:2020ynm}
(see also \cite{Jegerlehner:2017gek,Colangelo:2018mtw,Hoferichter:2019mqg,Davier:2019can,Keshavarzi:2019abf,Hoid:2020xjs,Czarnecki:2002nt, Melnikov:2003xd,Aoyama:2012wk, Gnendiger:2013pva})
\footnote{ There are arguments  
	on the precise value of the SM prediction $a^{\rm SM}_\mu$, for example,
	in Ref.~\cite{Borsanyi:2020mff}.
}.
Therefore, this experimental result has stimulated an interest in new physics contributions to this observable.  Indeed, comprehensive  analyses 
of the electric and  magnetic dipole moments  of leptons 
are given in  the SM Effective Field Theory (SMEFT), i.e., under the hypothesis of new degrees of freedom
above the electroweak scale \cite{Panico:2018hal,Aebischer:2021uvt,Allwicher:2021rtd,Kley:2021yhn}.
The phenomenological discussion of NP has appeared in the light of  the anomaly of the muon $(g-2)_\mu$ in the SMEFT
\cite{Isidori:2021gqe}.

Several years after the pioneering analysis in Ref.\cite{Buchmuller:1985jz}, 
the first complete list of the non-redundant SMEFT Lagrangian terms up to dimension-six has been presented in~Ref.\cite{Grzadkowski:2010es}.
When all the possible flavor structures are taken into account in the absence of any flavor symmetry, 
a large proliferation in the number of independent coefficients in the SMEFT occurs; there are 1350 CP-even and 1149 CP-odd independent coefficients for the dimension-six operators~\cite{Alonso:2013hga}.  
The flavor symmetry  reduces the number of independent parameters of the flavor sector.  
Above all, the flavor symmetries $U(3)^5$ and $U(2)^5$ have been successfully 
applied to the SMEFT~\cite{Faroughy:2020ina}.
The $U(3)^5$ flavor symmetry is the maximal flavor symmetry allowed by the SM gauge sector, 
while $U(2)^5$ is the corresponding  subgroup acting only on the first two (light) families \cite{Barbieri:2011ci,Barbieri:2012uh,Blankenburg:2012nx}. 
The $U(3)^5$ allows us to apply the Minimal Flavor Violation (MFV) hypothesis~\cite{Chivukula:1987py,DAmbrosio:2002vsn}, 
which is the most restrictive hypothesis consistent with the SMEFT,
and suppress non-standard contributions to flavor-violating observables~\cite{DAmbrosio:2002vsn}. 

On the other hand, we have already discussed the SMEFT with the modular flavor symmetry \cite{Kobayashi:2021uam,Kobayashi:2021pav}. 
Indeed, the well-known finite groups $S_3$, $A_4$, $S_4$ and $A_5$
are isomorphic to the finite modular groups 
$\Gamma_N$ for $N=2,3,4,5$, respectively\cite{deAdelhartToorop:2011re}.
The lepton mass matrices have been given successfully  in terms of {$\Gamma_3\simeq A_4$} modular forms \cite{Feruglio:2017spp}.
Modular invariant flavor models have also been proposed on the $\Gamma_2\simeq  S_3$ \cite{Kobayashi:2018vbk},
$\Gamma_4 \simeq  S_4$ \cite{Penedo:2018nmg} and  
$\Gamma_5 \simeq  A_5$ \cite{Novichkov:2018nkm,Ding:2019xna}.
Other finite groups are also derived from magnetized D-brane models \cite{Kobayashi:2018bff}.
By using these modular forms, the flavor mixing of quarks and leptons has been discussed successfully in these years
since the non-Abelian finite groups are long familiar
in  quarks and leptons
\cite{Altarelli:2010gt,Ishimori:2010au,Ishimori:2012zz,Kobayashi:2022moq,Hernandez:2012ra,King:2013eh,King:2014nza,Tanimoto:2015nfa,King:2017guk,Petcov:2017ggy,Feruglio:2019ktm}.

In the modular invariant theories of the finite group, the quark and lepton mass matrices are written in terms of modular forms, which are holomorphic functions of the modulus $\tau$.
The arbitrary symmetry breaking sector of the conventional models based on  flavor  symmetries is replaced by the moduli space, and then Yukawa couplings are given by modular forms.  
Phenomenological studies of the lepton flavors have been done
based on  $A_4$ \cite{Criado:2018thu,Kobayashi:2018scp,Ding:2019zxk}, 
$S_4$ \cite{Novichkov:2018ovf,Kobayashi:2019mna,Wang:2019ovr} and 
$ A_5$ \cite{Novichkov:2018nkm,Ding:2019xna}.
A clear prediction of the neutrino mixing angles and the Dirac CP phase was given in  the  simple lepton mass matrices with
the $A_4$ modular symmetry \cite{Kobayashi:2018scp}.
The Double Covering groups  $\rm T'$~\cite{Liu:2019khw,Chen:2020udk}
and $S_4'$ \cite{Novichkov:2020eep,Liu:2020akv} were also
realized in the modular symmetry.
Furthermore, phenomenological studies have been developed  in many works
\cite{deMedeirosVarzielas:2019cyj,
	Asaka:2019vev,Ding:2020msi,Asaka:2020tmo,Behera:2020sfe,Mishra:2020gxg,deAnda:2018ecu,Kobayashi:2019rzp,Novichkov:2018yse,Kobayashi:2018wkl,Okada:2018yrn,Okada:2019uoy,Nomura:2019jxj, Okada:2019xqk,
	Kariyazono:2019ehj,Nomura:2019yft,Okada:2019lzv,Nomura:2019lnr,Criado:2019tzk,
	King:2019vhv,Gui-JunDing:2019wap,deMedeirosVarzielas:2020kji,Zhang:2019ngf,Nomura:2019xsb,Kobayashi:2019gtp,Lu:2019vgm,Wang:2019xbo,King:2020qaj,Abbas:2020qzc,Okada:2020oxh,Okada:2020dmb,Ding:2020yen,Nomura:2020opk,Nomura:2020cog,Okada:2020rjb,Okada:2020ukr,Nagao:2020azf,Nagao:2020snm,Yao:2020zml,Wang:2020lxk,Abbas:2020vuy,
	Okada:2020brs,Yao:2020qyy,Feruglio:2021dte,King:2021fhl,Chen:2021zty,Novichkov:2021evw,Du:2020ylx,Kobayashi:2021jqu,Ding:2021zbg,Kuranaga:2021ujd,Li:2021buv,Tanimoto:2021ehw,Okada:2021aoi,Kobayashi:2021ajl,Dasgupta:2021ggp,Nomura:2021ewm,Nagao:2021rio,Nomura:2021yjb,Nomura:2021aep,Okada:2021qdf,Ding:2021eva,Qu:2021jdy,Zhang:2021olk,Wang:2021mkw,
	Wang:2020dbp,Ko:2021lpx,Nomura:2021pld,Nomura:2022hxs,Otsuka:2022rak,Ding:2021iqp,Charalampous:2021gmf,Liu:2021gwa,Novichkov:2022wvg,Kikuchi:2022txy
}.


In this paper, we focus on the lepton flavor violation (LFV), lepton $(g-2)_{\mu,\,e}$ and the electric dipole moment (EDM) of the electron.
Since we make the assumption that NP is heavy and can be given by the
SMEFT Lagrangian,
we discuss  the dipole operators of leptons and their Wilson coefficients  at the electroweak scale.
As the flavor symmetry, we  take the level 3 finite modular group, 
$\Gamma_3$  since  the property of $A_4$ flavor symmetry has been well known
\cite{Ma:2001dn,Babu:2002dz,Altarelli:2005yp,Altarelli:2005yx,
	Shimizu:2011xg,Petcov:2018snn,Kang:2018txu}. 
{Although modular flavor models have been constructed in the supersymmetric framework so far, the modular invariant SMEFT will be 
	realized in the so-called moduli-mediated supersymmetry breaking scenario. Indeed, the soft supersymmetry breaking terms are generated 
	in a modular invariant way \cite{Kikuchi:2022pkd}. 
	Furthermore, higher-dimensional operators also keep the modular invariance in a certain class of the string EFT 
	in which $n$-point couplings of matter fields are written by a product of 3-point couplings. We adopt this Ansatz (called stringy Ansatz) to constrain the higher-dimensional operators in the SMEFT. This meets the MFV hypothesis \cite{Kobayashi:2021uam}.}	
Based on the tensor decomposition of $A_4$ modular group, we discuss the bilinear  operators of leptons  at nearby fixed point $\tau=i$,
where observed  lepton masses and mixing angles are well reproduced.

The paper is organized as follows.
In section \ref{sec:string}, we present our framework, that is the stringy Ansatz.
In section \ref{sec:WC}, we discuss the flavor structure of the Wilson coefficients
of the leptonic dipole operator in mass basis.
In section \ref{sec:pheno}, we discuss the phenomenology of $(g-2)_{e,\mu}$,
LFV and the electron EDM.
Section \ref{sec:summary} is devoted to the summary.
In Appendix \ref{sect:exp}, we present the experimental constraints 
on the leptonic dipole operators.
In Appendix \ref{sec:mod}, we summarize $A_4$ modular symmetry briefly.
In Appendix \ref{massmatrixmodel}, we present a flavor model in $A_4$ modular symmetry.
In Appendix \ref{massmatrix-i}, we show the charged lepton mass matrix explicitly 
at the nearby $\tau=i$.

\section{Stringy Ansatz}
\label{sec:string}


In this section, we comment on an ultraviolet origin of the SMEFT with the $\Gamma_N$ modular flavor symmetry, {following Refs. \cite{Kobayashi:2021uam,Kobayashi:2021pav}.} 
	In particular, we focus on the superstring theory for physics beyond the SM. 
	It was known that $n$-point couplings $y^{(n)}$ of matter fields are written by products of 3-point couplings $y^{(3)}$, 
	i.e., $y^{(n)} = (y^{(3)})^{n-2}$ in a certain class of string compactifications. 
	For instance, 4-point couplings $y^{(4)}_{ijk\ell}$ of matter fields are given by
	\begin{equation}
	\label{eq:Ansatz}
	y^{(4)}_{ijk\ell}=\sum_m y^{(3)}_{ijm}y^{(3)}_{mk\ell}\,,
	\end{equation}
	up to an overall factor. 
	Here, the virtual modes $m$ are light or heavy modes, depending on the compactifications. 
	It indicates that the flavor structure of 3-point couplings and higher-dimensional operators has a common 
	origin in string compactifications. 
	Since the flavor symmetry in 3-point couplings controls the structure of four-dimensional EFT, 
	it satisfies the criterion of the MFV hypothesis \cite{Kobayashi:2021uam}. 
	As pointed out in Ref. \cite{Kobayashi:2021uam}, such a relation holds at the low-energy scale below the 
	compactification scale. 

{In the string-derived EFT, these $n$-point couplings depend on moduli fields, reflecting the geometric 
	symmetry of compact six-dimensional space. 
	For instance, the $SL(2,\mathbb{Z})$ modular symmetry is the geometric symmetry of the torus $T^2$ as well as 
	the orbifold $T^2/\mathbb{Z}_2$\footnote{The modular symmetries on higher-dimensional toroidal orbifolds were also discussed in Ref. \cite{Kobayashi:2020hoc}.}. 
	Recalling that the transformation of matter zero-modes on toroidal backgrounds is also given by the finite subgroup of the 
	modular symmetry (see, e.g., for heterotic string theory~\cite{Ferrara:1989qb,Lerche:1989cs,Lauer:1990tm} and for magnetized brane models~\cite{Kobayashi:2018rad,Ohki:2020bpo,Kikuchi:2020frp,Kikuchi:2020nxn,Kikuchi:2021ogn,Almumin:2021fbk})\footnote{See also \cite{Baur:2019iai,Nilles:2020kgo,Baur:2020jwc,Nilles:2020gvu}.}, 
	the flavor symmetry of matter zero-modes is determined by the modular symmetry in the low-energy effective action. 
	Furthermore, the modular symmetry restricts the form of $n$-point couplings in a modular symmetric way. 
	Much larger symplectic modular symmetries are possible in Calabi-Yau compactifications \cite{Strominger:1990pd,Candelas:1990pi} 
	whose phenomenological aspects were studied in Refs. \cite{Ishiguro:2020nuf,Ishiguro:2021ccl}.
	As a result, the flavor structure of Yukawa couplings and higher-dimensional operators are controlled by the modular flavor symmetry in the various class of string compactifications. 
	Note that the supersymmetry breaking sector also respects the flavor symmetry as seen in the soft supersymmetry 
	breaking terms induced by the moduli fields \cite{Kikuchi:2022pkd}.}

{Let us ignore the dynamics of moduli fields, meaning that moduli-dependent couplings are considered spurions 
	under the modular symmetry. 
	Then, the modular symmetry plays an important role in the concept of the MFV. In the original MFV scenario, 
	Yukawa couplings behave as $({\bf 3},{\bf \bar 3},1,1,1)$, $({\bf 3},1,{\bf \bar 3},1,1)$, and $(1,1,1,{\bf 3},{\bf \bar 3})$ in 
	the $U(3)^5 = U(3)_Q \otimes U(3)_U \otimes U(3)_D \otimes U(3)_L \otimes U(3)_E$  flavor symmetry. 
	On the other hand, in the string EFT at the leading order, $U(2)^5$ flavor symmetry is realized due to the rank 1 Yukawa couplings of matter fields~\cite{Ibanez:2012zz}. 
	It is interesting to analyze the phenomenological aspects of string-derived low-energy effective action with some modular symmetries which would be realized in toroidal as well as Calabi-Yau compactifications. 
	Indeed, the modular symmetry and the Ansatz Eq.(\ref{eq:Ansatz}) are powerful to predict the leptonic phenomena of flavors, as will be discussed in the next section. 
	In this paper, for concreteness, we study the SMEFT with the level 3 finite modular group $\Gamma_3$ for the flavor symmetry by imposing 
	the stringy Ansatz Eq.(\ref{eq:Ansatz}) on the higher-dimensional operators. 
	Remarkably, the lepton masses and mixing angles are well fitted with the observed data when the modulus field $\tau$ is close to the fixed point 
	$\tau= i$ in the $SL(2,\mathbb{Z})$ moduli space. 
	In subsequent sections, we discuss the higher-dimensional operators relevant to the lepton sector in more detail.}

\section{Wilson Coefficients of dipole operator in mass basis}
\label{sec:WC}

We take the assumption that NP is heavy and can be given by
the {SMEFT} Lagrangian.
Let us focus on  the dipole operators of leptons and their Wilson coefficients  at the weak scale as:
\begin{align}
&\mathcal{O}_{\underset{LR}{e\gamma}}
= \frac{v }{\sqrt{2}}  \overline{E}_{L}  \sigma^{\mu\nu} E_{R} F_{\mu\nu}\,,\qquad\qquad
	\CL{e\gamma}{LR}^\prime=
\begin{pmatrix}
	\CL{e\gamma}{ee}^\prime &\CL{e\gamma}{e\mu}^\prime
	 &\CL{e\gamma}{e\tau}^\prime\\
	\CL{e\gamma}{\mu e}^\prime &\CL{e\gamma}{\mu\mu}^\prime
&\CL{e\gamma}{\mu\tau}^\prime\\
	\CL{e\gamma}{\tau e}^\prime &\CL{e\gamma}{\tau\mu}^\prime
&\CL{e\gamma}{\tau\tau}^\prime\\
\end{pmatrix} 
\,,
\nonumber\\
&
\mathcal{O}_{\underset{RL}{e\gamma}}
= \frac{v }{\sqrt{2}}  \overline{E}_{R}  \sigma^{\mu\nu} E_{L} F_{\mu\nu}\,,\qquad\qquad
\CL{e\gamma}{RL}^\prime=\CL{e\gamma}{LR}^{\prime\, \dagger}
\,,
\label{dipole-operators}
\end{align}
where $E_L$ and $E_R$ denote three flavors of left-handed and
right-handed leptons, respectively,
and $v$ denotes the vacuum expectation value (VEV) of the Higgs field $H$.
Here the prime of the Wilson coefficient indicates the {flavor} basis corresponding to the mass-eigenstate basis of charged leptons.
The relevant effective Lagrangian is written as:
\begin{align}
\mathcal{L}_{\rm dipole}=
\frac{1}{\Lambda^2}\,\left (
\CL{e\gamma}{LR}^\prime\mathcal{O}_{\underset{LR}{e\gamma}}
+\CL{e\gamma}{RL}^\prime\mathcal{O}_{\underset{RL}{e\gamma}}
\right )
\,,
\end{align}
where $\Lambda$ is a certain mass scale of NP  in the effective theory.

In the following discussions, we take the $A_4$ modular symmetry for leptons.
Most of modular flavor models are supersymmetric models.
Since we study the model below the supersymmetry breaking scale, 
the light modes are exactly the same as the SM with two doublet Higgs models.
{Note that the modular symmetry is still a symmetry of the low-energy 
effective action below the supersymmetry breaking scale, 
as confirmed in the moduli-mediated supersymmetry breaking scenario.}

\subsection{Representation of  charged leptons in $A_4$ modular invariant model}

We take a  simple $A_4$ modular-invariant flavor model of leptons,
which is successful in reproducing neutrino masses and mixing angles, as shown explicitly in
  Appendix \ref{massmatrixmodel}.
In the model, the left-handed charged leptons compose a $A_4$ triplet
 $\bf  3$ 
and the three right-handed ones are  $A_4$ three different singlets.
Then, those are expressed as follows:

\begin{align}
\begin{aligned}  E_L=
\begin{pmatrix}
e_L  \\
\mu_L \\
\tau_L
\end{pmatrix}\, ,
\quad 
\bar E_L=
\begin{pmatrix}
\bar e_L  \\
\bar \tau_L \\
\bar \mu_L
\end{pmatrix}\, , \quad 
(e^c_R,\,\mu^c_R,\,\tau^c_R)=(1,\, 1'',\, 1')\,, \quad
( e_R,\, \mu_R,\,\tau_R)=(1,\, 1',\, 1'')\,.
\end{aligned}
\end{align}
It is noticed that leptons of second and third families 
are exchanged each other in $\bar E_L$.
As seen in the Table of  Appendix \ref{massmatrixmodel}, 
{
both $E_L$ and  $\bar E_L$ have the same modular weight,  $-k=-2$.}
On the other hand, $k=0$ for  $e^c_R$, $e_R$, etc..

The holomorphic and anti-holomorphic modular forms {of} weight 2 
compose the $A_4$ triplet: 
\begin{align}
&\begin{aligned}
Y(\tau)=
\begin{pmatrix}
Y_1(\tau)  \\
Y_2(\tau) \\
Y_3(\tau)
\end{pmatrix}\,,\qquad 
\overline {Y(\tau)}\equiv Y^{*}(\tau)=
\begin{pmatrix}
Y_1^*(\tau)  \\
Y_3^*(\tau) \\
Y_2^*(\tau)
\end{pmatrix}\,,
\end{aligned}
\end{align}
where modular forms are given explicitly in Appendix \ref{modular-form}.


\subsection{$[\,\bar E_R \Gamma  E_L\,]$ and $[\,\bar E_L \Gamma  E_R\,]$ bilinears in the flavor space}
\label{bilinear-flavor}	

In order to investigate  the flavor structure of the  Wilson coefficient of the dipole operator,
let us begin with discussing the  holomorphic operator
of charged leptons, 
$[\,\bar E_R \Gamma  E_L\,]$ and
anti-holomorphic operator $[\,\bar E_L\Gamma E_R\,]$ in the flavor space,
where $\Gamma$ denotes the relevant Lorentz structure.
The magnitudes of  $LR$ couplings are proportional to modular forms.
Taking account of $\bar E_R=(e^c,\mu^c,\tau^c)$, we can decompose  the operator
 in terms of  the $A_4$ triplet  holomorphic and anti-holomorphic modular forms {of} weight 2 
   in the basis of Eq.\,\eqref{ST} for $S$ and $T$.
   The result has been given in Ref.\cite{Kobayashi:2021pav} as:
\begin{align}
[\,\bar E_R\Gamma  E_L\,]
&\Rightarrow  [\,\bar E_R \Gamma Y({\tau}) E_L\,]_{\bf 1}
= \nonumber\\
& [\,{\alpha_{e}\,  \bar e_R }\Gamma
(Y_{1}({\tau}) e_L+Y_{2}({\tau}) \tau_L
+Y_{3}({\tau}) \mu_L)_1+{\beta_{e}\, \bar \mu_R}\Gamma
(Y_{3}({\tau}) \tau_L+Y_{1}({\tau}) \mu_L +Y_{2}({\tau}) e_L)_{\bf 1'}
\nonumber \\
&+ {\gamma_{e}\,  \bar \tau_R} \Gamma  (Y_{2}({\tau}) \mu_L +Y_{1}({\tau}) \tau_L +Y_{3}({\tau}) e_L)_{\bf 1''}]
\,,\nonumber\\
[\,\bar E_L  \Gamma E_R\,]
&\Rightarrow [\,\bar E_L   Y^*({\tau})\Gamma E_R\,]_{\bf 1}
= \nonumber\\
&[\,{ \alpha_{e}^*\,   e_R } \Gamma (Y_{1}^*({\tau})\bar e_L+Y_{2}^*({\tau})\bar \tau_L
+Y_{3}^*({\tau}) \bar \mu_L)_{\bf 1}+ 
{ \beta_{e}^*\, \mu_R } \Gamma
(Y_{3}^*({\tau}) \bar \tau_L+Y_{1}^*({\tau}) \bar{\mu}_L +Y_{2}^*({\tau}) \bar d_L)_{\bf 1''}   \nonumber\\
&+{ \gamma_{e}^*\,  \tau_R} \Gamma 
(Y_{2}^*({\tau}) \bar{\mu}_L+Y_{1}^*({\tau}) \bar{\tau}_L +Y_{3}^*({\tau})\bar e_L)_{\bf 1'}]
\,,
\label{DLR}
\end{align}	
where 
the subscript $1$, $1'$, $1''$ denote three $A_4$ singlets, respectively.
The parameters	$\alpha_{e}$,
$\beta_{e}$ and	$\gamma_{e}$ are constants.   

These expressions are written in the matrix representation as 
\begin{align} 
[\,\bar E_R \Gamma Y({\tau}) E_L\,]_{\bf 1} 
&= ( \bar e_R,  \bar \mu_R,  \bar \tau_R)\Gamma
\begin{pmatrix}
\alpha_e & 0 & 0 \\
0 & \beta_e & 0\\
0 & 0 & \gamma_e
\end{pmatrix}
\begin{pmatrix}
Y_1(\tau) & Y_3(\tau)& Y_2(\tau)\\
Y_2(\tau) & Y_1(\tau) &  Y_3(\tau) \\
Y_3(\tau) &  Y_2(\tau)&  Y_1(\tau)
\end{pmatrix}
\begin{pmatrix}
e_L\\
\mu_L\\
\tau_L
\end{pmatrix}
\,, 
\label{RLmatrix}\\ %
[\,\bar E_L   Y^*({\tau})\Gamma E_R\,]_{\bf 1} 
&= (\bar e_L, \bar \mu_L, \bar \tau_L)\Gamma
\begin{pmatrix}
Y^*_1(\tau) & Y^*_2(\tau)& Y^*_3(\tau)\\
Y^*_3(\tau) & Y^*_1(\tau) &  Y^*_2(\tau) \\
Y^*_2(\tau) &  Y^*_3(\tau)&  Y^*_1(\tau)
\end{pmatrix}
\begin{pmatrix}
\alpha_e^* & 0 & 0 \\
0 & \beta_e^* & 0\\
0 & 0 & \gamma_e^*
\end{pmatrix}
\begin{pmatrix}
e_R\\
\mu_R\\
\tau_R
\end{pmatrix}
\,.
\label{LRmatrix}
\end{align}  
It is useful to compare them with  the charged lepton mass matrix $M_E$
in Appendix \ref{massmatrixmodel}.
The mass matrix is given in terms of weight 2 modular forms as:
\begin{align} 
M_E&= v_d
\begin{pmatrix}
\alpha_{e(m)} & 0 & 0 \\
0 & \beta_{e(m)} & 0\\
0 & 0 &\gamma_{e(m)}
\end{pmatrix}
\begin{pmatrix}
Y_1(\tau) & Y_3(\tau)& Y_2(\tau)\\
Y_2(\tau) & Y_1(\tau) &  Y_3(\tau) \\
Y_3(\tau) &  Y_2(\tau)&  Y_1(\tau)
\end{pmatrix}_{RL}\,, 
\label{charged-massmatrix} 
\end{align} 
where 
the VEV of the Higgs field $H_d$ is denoted by $ v_d$.
Parameters $\alpha_{e(m)}$, $\beta_{e(m)}$,  $\gamma_{e(m)}$ can be taken to be  real constants.
Since the bilinear operators appear in four-field operators
corresponding to Eq.\eqref{dipole-operators} by replacing $v$ with $H$,
it is reasonable to assume 
\begin{align}
\label{eq:alpha-relation0}
\alpha_e=\kappa\alpha_{e(m)}, \qquad \beta_e=\kappa\beta_{e(m)}, \qquad \gamma_e=\kappa\gamma_{e(m)},
\end{align}
from the viewpoint of the Ansatz Eq.\,(\ref{eq:Ansatz}),  
where the mode $m$ may correspond to $H_d$.
Here, $\kappa$ is a common constant.
Hereafter, we set $\kappa=1$ by taking the relevant normalization. 
In this case, the matrix structure of bilinear operators $[\,\bar E_R \Gamma  E_L\,]$  is 
exactly the same as the mass matrix.
Obviously, the bilinear operator matrix is diagonal in the basis for mass eigenstates.
The {flavor changing (FC)} processes such as $\mu \to e$, $\tau \to \mu$
and  $\tau \to e$ never happen. 
Hence, we obtain the very clear results in the modular symmetric SMEFT with the Ansatz Eq.\,(\ref{eq:Ansatz}).

However, additional unknown modes  in Eq.\,(\ref{eq:Ansatz}) may cause
the flavor violation.
If the relation of Eq.\eqref{eq:alpha-relation0} is violated, the situation would change drastically.
Consider the case  {that} such violations are small  in the following discussions:
\begin{align}
\label{eq:alpha-relation}
\alpha_e-\alpha_{e(m)} \ll \alpha_e\,, \qquad \beta_e-\beta_{e(m)} \ll \beta_e\,, \qquad \gamma_e-\gamma_{e(m)} \ll \gamma_e\,.
\end{align}

We summarize the $A_4$ flavor  coefficients 
of Eqs.\eqref{RLmatrix} and \eqref{LRmatrix}
  in Table \ref{LR-Table} for  relevant  bilinear  operators
of  charged leptons \cite{Kobayashi:2021pav}, where 
the overall strength of the NP effect is not included
\footnote{The overall strength of the NP effect
	is also omitted in coefficients of other Tables.}
.
\begin{table}[h]
	\small
	\centering
	\begin{tabular}{|c|c|c|c|c|c|c|} \hline 
		\rule[14pt]{0pt}{2pt}   		
		$\begin{matrix}
		{\bar R L} \ {\rm operators} \\ {\bar LR}\ {\rm operators}
		\end{matrix} $ 
		&
		$\begin{matrix}
		\bar \mu_R \Gamma  \tau_L^{}\\\bar \mu_L \Gamma  \tau_R^{}
		\end{matrix} $ 
		& 	$\begin{matrix}\bar e_R \Gamma  \tau_L^{}\\\bar e_L \Gamma  \tau_R^{}
		\end{matrix} $ 
		&	$\begin{matrix}\bar e_R \Gamma  \mu_L^{}\\\bar e_L \Gamma \mu_R^{}
		\end{matrix} $ 
		& 	$\begin{matrix}\bar e_R \Gamma  e_L^{}\\\bar e_L \Gamma  e_R^{}
		\end{matrix} $ 
		&	$\begin{matrix}\bar \mu_R \Gamma  \mu_L^{}\\\bar \mu_L \Gamma \mu_R^{}
		\end{matrix} $ 
		&$\begin{matrix}\bar\tau_R \Gamma \tau_L^{}\\\bar \tau_L \Gamma \tau_R^{}
		\end{matrix} $ 
		\\  \hline
		\rule[14pt]{0pt}{1pt} 			
		$\begin{matrix}
		{\rm Flavor\  Coefficients} 
		\end{matrix} $ 
		&
		$\begin{matrix}	{\beta_{e}\,  Y_{3}({\tau_e}) }\\
		{\gamma_{e}\,  Y_{2}^*({\tau_e}) }
		\end{matrix} $ 
		& 	$\begin{matrix}	{ \alpha_{e}\,  Y_{2}({\tau_e}) }\\
		{\gamma_{e}\,  Y_{3}^*({\tau_e}) }
		\end{matrix} $  &
		$\begin{matrix}{ \alpha_{e}\,  Y_{3}({\tau_e}) }\\
		{\beta_{e}\,  Y_{2}^*({\tau_e}) }
		\end{matrix} $ 
		& 	$\begin{matrix}	{ \alpha_{e}\,  Y_{1}({\tau_e}) }\\
		{\alpha_{e}\,  Y_{1}^*({\tau_e}) }
		\end{matrix} $  &
		$\begin{matrix}{ \beta_{e}\,  Y_{1}({\tau_e}) }\\
		{\beta_{e}\,  Y_{1}^*({\tau_e}) }
		\end{matrix} $
		 &
		$\begin{matrix}{ \gamma_{e}\,  Y_{1}({\tau_e}) }\\
		{\gamma_{e}\,  Y_{1}^*({\tau_e}) }
		\end{matrix} $  \\ \hline
	\end{tabular}
	\caption{Flavor coefficients of  the  bilinear operators of  charged leptons in the flavor basis.}
	\label{LR-Table}
	\normalsize
\end{table}


It is noticed that the above operators are given in the flavor basis.
In order to obtain the mass eigenstate of leptons,
we must fix the modulus $\tau$.
The value of $\tau$ depends on models.
In the  model  in Appendix \ref{massmatrixmodel},
 $\tau$ is close to the fixed point of the modulus,  $\tau=i$.

\subsubsection{Mass eigenstate at nearby $\tau=i$}
At the fixed point $\tau=i$, the flavor structure is too simple
 to reproduce the observed lepton mixing angles.
The modulus $\tau$ is deviated from the fixed points $\tau=i$
in order to get the observed  lepton  masses and mixing angles.
Indeed, the successful charged lepton mass matrix
has been obtained  at nearby $\tau=i$ \cite{Okada:2020brs}.
By using a small dimensionless parameter $\epsilon$,
we put the modulus value as  $\tau=i+\epsilon$.
Then, approximate behaviors of the ratios of modular forms
are \cite{Okada:2020ukr}: 
\begin{align}
\begin{aligned}
\frac{Y_2(\tau)}{Y_1(\tau)}\simeq (1+\epsilon_1)\, (1-\sqrt{3}) \, , \quad 
\frac{Y_3(\tau)}{Y_1(\tau)}\simeq (1+\epsilon_2)\, (-2+\sqrt{3}) \, ,
\quad \epsilon_1=\frac{1}{2} \epsilon_2\simeq 2.05\,i\,\epsilon\,.
\end{aligned}
\label{epS12}
\end{align}
These approximate  forms are  agreement with exact numerical values within  $0.1\,\%$
for $|\epsilon|\leq 0.05$.

	The charged lepton mass matrix is diagonalized 
	by the following transformation which is also  shown in Appendix \ref{massmatrix-i}:	
	\begin{align}
	&E_L \rightarrow  E_L^m\equiv U_{Lme}^\dagger  U_{S}\, E_L\,,
	\qquad\qquad \  \,
	\bar E_L \rightarrow \bar E_L^m \equiv 
	{\bar E}_L\,U_{S}^T  U_{Lme} \,,
	\nonumber\\
	&E_R \rightarrow  E_R^m\equiv U_{Rme}^\dagger\, U_{12}^T\, E_R\,, 
	\qquad\quad\quad
	\bar E_R \rightarrow \bar E_R^m \equiv 
	{\bar E}_R\, U_{12}\,U_{Rme}  \,,
	\label{massbase-Se}
	\end{align}	
	where
	\begin{align}
	U_S =
	\frac{1}{2\sqrt{3}}
	\begin{pmatrix}
	2  & 2 & 2\\
	\sqrt{3}+1  &  -2 & \sqrt{3}-1\\
	\sqrt{3}-1 & -2  & \sqrt{3}+1
	\end{pmatrix}\,, \qquad 	U_{12}=
	\begin{pmatrix}
		0 & 1 & 0\\  1& 0 & 0 \\ 0 & 0 &1
	\end{pmatrix}\,.
	\end{align}
	The mixing matrices are parametrized as:
	\begin{align}
	U_{Lme}\simeq  P_e^*
	\begin{pmatrix}
	1 & s_{L12}^{e}  &s_{L13}^{e}\\
	-s_{L12}^{e} & 1 & s_{L23}^{e}\\
	s_{L12}^{e} s_{L23}^{e} -s_{L13}^{e}   & -s_{L23}^{e}  & 1
	\end{pmatrix}
	,
	\quad 
	U_{Rme}\simeq 
	\begin{pmatrix}
	1 & s_{R12}^{e} & s_{R13}^{e}\\
	-s_{R12}^{e} & 1 & s_{R23}^{e} \\
	s_{R12}^e s_{R23}^e-s_{R13}^{e} & -s_{R23}^{e} & 1
	\end{pmatrix}
	\label{ULRe}\,,
	\end{align}
	where the phase matrix $P_e$ is
	\begin{align}
	P_e=
	\begin{pmatrix}
	e^{ i\eta_e  }& 0 & 0 \\
	0&1 & 0\\
	0 & 0 & 1 \\
	\end{pmatrix} \,, \qquad \eta_e=\arg \epsilon_1 \,.
	\end{align}
	The mixing angles are given 
	as seen in Appendix \ref{massmatrix-i}:
	\begin{align}
	&s^e_{L12}\simeq -|\epsilon_1^*|\,,\qquad\quad\   
	s^e_{L23}\simeq -\frac{\sqrt{3}}{4}\frac{\tilde\alpha_{e(m)}^2}{\tilde\gamma_{e(m)}^2},
	\quad\qquad\ \quad  s^e_{L13}\simeq -\frac{\sqrt{3}}{3}|\epsilon_1^*|\,,
	\nonumber\\
	&s^e_{R12}\simeq -\frac{\tilde\beta_{e(m)}}{\tilde\alpha_{e(m)}}\,,
	\qquad\   s^e_{R23}\simeq-\frac12\frac{\tilde\alpha_{e(m)}}{\tilde\gamma_{e(m)}}\,,\qquad\qquad\quad  s^e_{R13}\simeq-\frac12\frac{\tilde\beta_{e(m)}}{\tilde\gamma_{e(m)}}\,,
	\label{mixing-S}
	\end{align}
	where  $\tilde \alpha_{e(m)}=(6-3\sqrt{3})  Y_1(i)\alpha_{e(m)}$,
	$\tilde \beta_{e(m)}= (6-3\sqrt{3})  Y_1(i) \beta_{e(m)}$ and 
	$\tilde \gamma_{e(m)} =(6-3\sqrt{3}) Y_1(i)  \gamma_{e(m)}$.
	Indeed, the numerical fit was succeeded 
	in the case of $\tilde\gamma_{e(m)}^2\gg \tilde\alpha_{e(m)}^2\gg\tilde\beta_{e(m)}^2$ \cite{Okada:2020brs}.
In the mass eigenstate,
the $A_4$ flavor coefficients of charged lepton bilinear operators 
are given in terms of mixing angles $s_{12}^e$, $s_{13}^e$
and $\epsilon_1$ at $\tau=i+\epsilon$ in Table \ref{Co-Se}
	\footnote{
		These results are different from ones in the previous our
		works \cite{Kobayashi:2021pav}. The previous result
		  was obtained  in  a flavor  basis of leptons where
	   the right-handed leptons are not rotated.
   However, 
   the previous results are also justified approximately
   due to  the different condition 
   from Eq.\eqref{eq:alpha-relation},
   such as $\alpha_e-\alpha_{e(m)} \sim \alpha_e$, etc..
}.

\begin{table}[h]
	\footnotesize{
			\begin{tabular}{|c|c|c|} \hline 
				\rule[14pt]{0pt}{2pt}   		
					$\begin{matrix}\bar \mu_R \Gamma  \tau_L^{}\\
				\bar \mu_L \Gamma  \tau_R^{}
				\end{matrix} $ 
				& 	$\begin{matrix}\bar e_R  \Gamma \tau_L^{}\\
				\bar e_L \Gamma  \tau_R^{}
				\end{matrix} $ 
				&	$\begin{matrix}\bar e_R \Gamma  \mu_L^{}\\
				\bar e_L \Gamma  \mu_R^{}
				\end{matrix} $ 
				\\  \hline
				\rule[14pt]{0pt}{2pt} 			
				\rule[14pt]{0pt}{1pt} 	
				$\begin{matrix}\frac{\sqrt{3}}{2}
				(\tilde\alpha_e+2s_{R23}^e \tilde \gamma_e)\\
				\rule[14pt]{0pt}{1pt} 	
				\hskip -0.3 cm (\sqrt{3}s_{23L}^e+s_{12L}^e|\epsilon_1^*|)
				\tilde \gamma_e-\frac32s_{R23}^e\tilde \alpha_e
				\end{matrix} $ 
				& 	$\begin{matrix}	\frac{\sqrt{3}}{2}
				(\tilde \beta_e-s_{12R}^e\tilde \alpha_e+
				2(s_{R13}^e-s_{R12}^es_{R23}^e)\tilde \gamma_e)\\
				\rule[14pt]{0pt}{1pt} 	
				(\sqrt{3}s_{13L}^e+|\epsilon_1^*|)\tilde \gamma_e
				\end{matrix} $  &
				$\begin{matrix}		\frac32 
				(\tilde \beta_e+s_{12R}^e\tilde \alpha_e)
				\\	
				\rule[14pt]{0pt}{1pt} 		\frac12  (3s_{12L}^e-\sqrt{3}s_{13L}^e+2|\epsilon_1^*|)
				\tilde \alpha_e
				\end{matrix} $ \\
				\hline 
			\end{tabular}
		}
	\caption{$A_4$ flavor coefficients of the FC lepton bilinear operators 
		at   $\tau=i+\epsilon$,
		{where $ {\cal O}(|\epsilon|^2)$ is neglected 
			because  the modular forms are expanded 
			in ${\cal O}(|\epsilon|)$, and 
			$\tilde \alpha_{e}=(6-3\sqrt{3})  Y_1(i)\alpha_{e}$,
			$\tilde \beta_{e}= (6-3\sqrt{3})  Y_1(i) \beta_{e}$, 
			$\tilde \gamma_{e} =(6-3\sqrt{3}) Y_1(i)  \gamma_{e}$.
		A common  overall factor  {$(1-\sqrt{3})$}
		is omitted in the coefficients. }
	}
	\label{Co-Se}
\end{table}

It is easily noticed that 
coefficients of $\bar \mu_L \Gamma  \tau_R$, $\bar e_L \Gamma  \tau_R$
and $\bar e_L \Gamma  \mu_R$ in Table \ref{Co-Se} are much suppressed in spite of 
$\alpha_e\not =\alpha_{e(m)}, \, \beta_e\not=\beta_{e(m)}, \, \gamma_e\not=\gamma_{e(m)}$,
 by inputting mixing angles of  Eq.\eqref{mixing-S} into them.
Indeed, we find that those coefficients are ${\cal O}(\epsilon_1\tilde\alpha^2_e/\tilde\gamma_e)$
for $\bar \mu_L \Gamma  \tau_R$ and  $\bar e_L \Gamma  \tau_R$
while 
 ${\cal O}(\tilde\beta^2_e/\tilde\alpha_e)$ for  $\bar e_L \Gamma  \mu_R$
 after calculations of the next-to-leading terms.
 Numerical values of these parameter are given
 at the best fit point as follows \cite{Okada:2020brs}:
 \begin{eqnarray}
 \tau=-0.080+ 1.007\, i\,,\ \  |\epsilon_1|=0.165\,,\ \ 
 \frac{\tilde \alpha_{e(m)}}{\tilde \gamma_{e(m)}}
 \simeq \frac{\tilde \alpha_{e}}{\tilde \gamma_{e}}=6.82\times 10^{-2}\,,
 \ \ 
 \frac{\tilde \beta_{e(m)}}{\tilde \alpha_{e(m)}}
 \simeq \frac{\tilde \beta_{e}}{\tilde \alpha_{e}}=1.50\times 10^{-2}\, .
 \nonumber\\ 
 \label{Lepton-model}
 \end{eqnarray}
 
However, the coefficients of the bilinear $\bar R L$ operators
 $\bar \mu_R \Gamma  \tau_L$, $\bar e_R \Gamma  \tau_L$ and 
  $\bar e_R \Gamma  \mu_L$ are not so suppressed.
  Those operators may lead to the sizable LFV decays.  
  Including 	a common  overall factor  {$(1-\sqrt{3})$},
  which  is omitted in Table \ref{Co-Se}, the coefficients are given as:
 \begin{eqnarray}
&& \CL{e\gamma}{\tau\mu }^\prime
=\frac{\sqrt{3}}{2}(1-\sqrt{3})\tilde\alpha_e\left ( 1-\frac{\tilde\gamma_e}{\tilde \gamma_{e(m)}} \frac{\tilde \alpha_{e(m)}}{\tilde\alpha_e}\right ) \,,\nonumber\\
&&\CL{e\gamma}{\tau e}^\prime =\frac{\sqrt{3}}{2}(1-\sqrt{3})\tilde\beta_e\left ( 1+\frac{\tilde\alpha_e}{\tilde \alpha_{e(m)}}\frac{\tilde \beta_{e(m)}}{\tilde\beta_e} 
-2\frac{\tilde\beta_{e(m)}}{\tilde\beta_e}\frac{\tilde\gamma_e}{\tilde \gamma_{e(m)}}  \right )\,,
 \nonumber\\
&& \CL{e\gamma}{\mu e}^\prime=\frac32(1-\sqrt{3})\tilde\beta_e
\left ( 1-\frac{\tilde\alpha_e}{\tilde \alpha_{e(m)}} 
\frac{\tilde \beta_{e(m)}}{\tilde\beta_e}\right ) \, ,
 \label{C-RL}
 \end{eqnarray}
 where ${\tilde\alpha_e}/{\tilde \alpha_{e(m)}}$
 ${\tilde\beta_e}/{\tilde \beta_{e(m)}}$
and  ${\tilde\gamma_e}/{\tilde \gamma_{e(m)}}$ are close to $1$
due to the condition of Eq.\eqref{eq:alpha-relation}.
 

On the other hand, the diagonal 
 coefficients of the bilinear $\bar R L$ operators
$\bar e_R \Gamma  e_L$,  $\bar \mu_R \Gamma  \mu_L$ and 
 $\bar \tau_R \Gamma  \tau_L$ are given as:
\begin{eqnarray}
  \CL{e\gamma}{e e}^\prime=3 \,(1-\sqrt{3})\tilde\beta_e
 |\epsilon_1^*| \,,\qquad
\CL{e\gamma}{\mu\mu}^\prime =\frac32\, (1-\sqrt{3})\tilde\alpha_e\,,
\qquad
\CL{e\gamma}{\tau\tau }^\prime
=\sqrt{3}\,(1-\sqrt{3})\tilde\gamma_e \, ,
\label{C-diagonal}
\end{eqnarray}
where the phase of $\epsilon_1$ is rotated away.
These give the anomalous magnetic moment of leptons.

\section{Phenomenology of $(g-2)_{\mu,\,e}$, LFV  and  EDM }
\label{sec:pheno}

The anomalous magnetic moment of the muon, $a_\mu =(g-2)_\mu/2$, is a  powerful probe 
 beyond the SM. The recent experimental measurement of $a_\mu$
by the E989 experiment at FNAL~\cite{Muong-2:2021ojo}, 
combined with the previous BNL result~\cite{Muong-2:2006rrc}, 
has indicated the discrepancy with the SM prediction reported 
in Ref.~\cite{Aoyama:2020ynm}.  
If this result is evidence of NP, we can relate it with other phenomena, 
   $(g-2)_{e}$, LFV processes and the {electron} EDM
  in the framework of the stringy Ansatz Eq.\,(\ref{eq:Ansatz}) with the modular symmetry.
We study the correlations among them in this section.


\subsection{$(g-2)_{\mu}$ and $(g-2)_{e}$}

 The NP of $(g-2)_{\mu}$ and $(g-2)_{e}$ {appears} in the diagonal
  {components} of the Wilson coefficient of the dipole operator
   at the mass basis.
 We have the ratios of the diagonal coefficients from
 Eq.\eqref{C-diagonal} as:
 \begin{eqnarray}
 \label{CG-ee}
&&\frac{ \CL{e\gamma}{e e}^\prime}{\CL{e\gamma}{\mu \mu}^\prime} =2\frac{\tilde\beta_e}{\tilde \alpha_{e}} |\epsilon_1^*|
\simeq 4.9\times 10^{-3} \,,
\qquad\qquad 
\frac{ \CL{e\gamma}{\mu\mu}^\prime}{\CL{e\gamma}{\tau \tau}^\prime} = \frac{\sqrt{3}}{2}\frac{\tilde\alpha_e}{\tilde \gamma_{e}}
\simeq 5.9\times 10^{-2} 
 \,,
\end{eqnarray}
where numerical values of Eq.\eqref{Lepton-model} are put
 for  $\tilde\beta_e/\tilde\alpha_e$, $\tilde\alpha_e/\tilde\gamma_e$ and $|\epsilon_1^*|$.
These predicted ratios are almost agree with the charged lepton mass ratios
$m_e/m_\mu=4.84\times 10^{-3}$ and $m_\mu/m_\tau=5.95\times 10^{-2}$.

If this  dipole operator is responsible for  the observed anomaly of
$(g-2)_{\mu}$, the magnitude of its Wilson coefficient  can be estimated
 as shown in Appendix \ref{sect:exp}.
By inputting the experimental value of Eq.\eqref{muon-data}, 
the real part of the Wilson coefficient of the muon $\CL{e\gamma}{\mu \mu}^\prime$
has been  {obtained} as seen  in Eq.\eqref{Cmumuexp} \cite{Isidori:2021gqe}.
Now, we can estimate the magnitude of  the electron $(g-2)_e$ anomaly 
by using the  relation in Eq.\eqref{CG-ee} as:
\begin{align}
\Delta a_{e} &= \frac{4 m_e}{e}   \frac{v}{\sqrt 2}\,\frac{1}{\Lambda^2} 
\text{Re} \,[\CL{e\gamma}{ee}^\prime] \simeq 5.8\times 10^{-14}\,,
\end{align}
where $\Lambda$  denotes a certain mass scale of NP.
It is easily seen that $\Delta a_{e}$ 
and  $\Delta a_{\mu}$ are proportional to the lepton {masses squared}.
This result is agreement with the naive scaling $\Delta a_\ell \propto m^2_\ell$\cite{Giudice:2012ms}.

In the electron {anomalous} magnetic moment, 
the  experiments \cite{Hanneke:2008tm} give
\begin{align}
a_e^\mathrm{Exp}=1\, 159\, 652\, 180.73(28)\times 10^{-12}\,,
\end{align}
while 
 the SM prediction crucially depends on the input value for the fine-structure constant $\alpha$.  Two latest determination
\cite{Parker:2018vye,Rb:2020} based on 
Cesium and Rubidium atomic recoils differ by more than $5\sigma$.
Those observations lead to the difference {from the} SM prediction
\begin{align}
&&\Delta a_e^{Cs} &= a_e^\mathrm{Exp} - a_e^\mathrm{SM,CS} = \brackets{-8.8 \pm 3.6} \times 10^{-13}~\,, \nonumber\\
&&\Delta a_e^{Rb} &= a_e^\mathrm{Exp} - a_e^\mathrm{SM,Rb} = \brackets{4.8 \pm 3.0} \times 10^{-13}~\,.
\end{align}
Our predicted value is small of one order
compared with the present observed one at present. 
We wait for the {precise} observation of the fine structure constant
  to test our framework.


\subsection{$(g-2)_{\mu}$ and $\mu\to e\gamma$}
The NP in the LFV  process
 is severely constrained by 
 the experimental  bound $\mathcal{B}\!\brackets{\mu^+ \to e^+ \gamma} < 4.2 \times 10^{-13}$ in the MEG experiment~\cite{TheMEG:2016wtm}.
We can discuss  the correlation 
between the anomaly of  the muon $(g-2)_\mu$ and
the LFV process $\mu\to e\gamma$ by using the Wilson coefficients
in Eqs.\eqref{C-RL} and  \eqref{C-diagonal}.
The  ratio is given as:
\begin{eqnarray}
\left |\frac{ \CL{e\gamma}{e \mu}^\prime}{\CL{e\gamma}{\mu \mu}^\prime}
\right |
=\frac{\tilde\beta_e}{\tilde \alpha_{e}}
\left | 1-\frac{\tilde\alpha_e}{\tilde \alpha_{e(m)}} 
\frac{\tilde \beta_{e(m)}}{\tilde\beta_e}\right |  \,.
\end{eqnarray}
Let us introduce small parameters $\delta_\alpha$,
$\delta_\beta$ and $\delta_\gamma$ as follows:
 \begin{eqnarray}
&& \frac{\tilde\beta_e}{\tilde \beta_{e(m)}}=\frac{\tilde\beta_{e(m)}+c_\beta}{\tilde \beta_{e(m)}}
=1+\frac{c_\beta}{\tilde \beta_{e(m)}}\equiv 1+\delta_\beta\, ,
\nonumber\\
&& \frac{\tilde\alpha_e}{\tilde \alpha_{e(m)}}=\frac{\tilde\alpha_{e(m)}+c_\alpha}{\tilde \alpha_{e(m)}}
=1+\frac{c_\alpha}{\tilde \alpha_{e(m)}}\equiv 1+\delta_\alpha\, ,
\nonumber\\ 
&& \frac{\tilde\gamma_e}{\tilde \gamma_{e(m)}}=\frac{\tilde\gamma_{e(m)}+c_\gamma}{\tilde \gamma_{e(m)}}
=1+\frac{c_\gamma}{\tilde \gamma_{e(m)}}\equiv 1+\delta_\gamma\, ,
\label{deviation}
\end{eqnarray}
where $c_\alpha$, $c_\beta$ and $c_\gamma$  are tiny  {contributions} from the  unknown mode of $m$ in Eq.\eqref{eq:Ansatz}. 
Putting the experimental  bound of this ratio in Eq.\eqref{eq:bound12} with Eq.\eqref{Lepton-model},
we obtain
\begin{eqnarray}
\left | 1-\frac{\tilde\alpha_e}{\tilde \alpha_{e(m)}} 
\frac{\tilde \beta_{e(m)}}{\tilde\beta_e}\right | 
\simeq |\delta_\beta -\delta_\alpha|< 1.4\times 10^{-3} \,,
\label{LFV-constraint}
\end{eqnarray}
which suggests
\begin{eqnarray}
\label{difference}
|\delta_\alpha| < {\cal O}(10^{-3}) \,,
\qquad |\delta_\beta|< {\cal O}(10^{-3}) \,,
\end{eqnarray}
without tuning between $\delta_\alpha$ and $\delta_\beta$.
Thus, additional contributions to $\tilde \alpha_{e(m)}$ and 
 $\tilde \beta_{e(m)}$ are at most ${\cal O}(10^{-3})$.
 
 
 It is emphasized that the NP signal of the $\mu\to e\gamma$ process comes from the  operator $\bar e_R \sigma_{\mu\nu}\mu_L$ mainly
 in our scheme.
 The angular distribution with respect to the muon polarization can distinguish between $\mu^+ \to e^+_L\gamma$ and $\mu^+ \to e^+_R\gamma$
 \cite{Okada:1999zk}.

 
 Let us consider the correlation among
 the LFV processes $\mu\to e\gamma$, $\tau\to \mu\gamma$
 and $\tau\to e\gamma$. 
 Since it depends on  $\delta_\alpha$,
 $\delta_\beta$ and $\delta_\gamma$,
 we consider two cases for these parameters.
 The first one is the case that  the additional unknown mode of $m$ is the Higgs-like mode,
 that is,  $\delta_\alpha\sim \delta_\beta\sim \delta_\gamma$. Then,
 we obtain ratios of the Wilson coefficients 
 by using  Eq.\eqref{C-RL} as:
 \begin{eqnarray}
 \frac{ \CL{e\gamma}{\tau e}^\prime}{\CL{e\gamma}{\mu e}^\prime} =\frac{1}{\sqrt{3}}\times {\cal O}(1)
 \,, \qquad\qquad
 \frac{ \CL{e\gamma}{\tau e}^\prime}{\CL{e\gamma}{\tau \mu}^\prime} =\frac{\tilde\beta_e}{\tilde \alpha_{e}}\times {\cal O}(1) \sim 10^{-2}\,,
 \label{case-1}
 \end{eqnarray}
 where the numerical value in Eq.\eqref{Lepton-model} is put.
 The decay rates are calculated in terms of Wilson coefficients as seen in Eq.\eqref{eq:Branching-ratio_lepton-decay}. In this case, we have 
 $\mathcal{B}(\tau \to \mu \gamma):
 \mathcal{B}(\tau \to e \gamma):
 \mathcal{B}(\mu \to e \gamma)\sim 10^4:1:10$,
  where we take account of the kinematical factor.
Since the present upper bounds of
$\mathcal{B}\!\brackets{\tau \to e \gamma}$
and $\mathcal{B}\!\brackets{\tau \to \mu \gamma}$ are 
 $3.3~ \times~ 10^{-8}$ and  $4.4~ \times~ 10^{-8}$ \cite{Zyla:2020zbs},
 respectively,  we expect the experimental test 
 of this prediction for $\tau\to \mu \gamma$ in the future.
 
 Another case is that    unknown mode of $m$ is the flavor blind one,
 that is  $c_\alpha=c_\beta=c_\gamma=c$ in Eq.\eqref{deviation}.
Therefore,  we have 
 $|\delta_\beta|\gg |\delta_\alpha|\gg |\delta_\gamma|$
 due to the hierarchy of  ${\tilde \beta_{e(m)}}\ll 
 {\tilde \alpha_{e(m)}}\ll {\tilde \gamma_{\gamma(m)}}$.
 We obtain the Wilson coefficients  by using Eq.\eqref{C-RL}:
  \begin{eqnarray}
&& \CL{e\gamma}{\tau \mu}^\prime
=\frac{\sqrt{3}}{2}(1-\sqrt{3})
{\tilde \alpha_e}\left (1-\frac{1+\delta_\gamma}{1+\delta_\alpha}
\right )
\simeq \frac{\sqrt{3}}{2}(1-\sqrt{3})
 {\tilde \alpha_e}\delta_\alpha\,, \nonumber\\
&&  \CL{e\gamma}{\tau e}^\prime
=\frac{\sqrt{3}}{2}(1-\sqrt{3})
{\tilde \beta_e}\left (1+\frac{1+\delta_\alpha}{1+\delta_\beta}
-2\frac{1+\delta_\gamma}{1+\delta_\beta}
\right )
\simeq \frac{\sqrt{3}}{2}(1-\sqrt{3})
{\tilde \beta_e}(\delta_\alpha+\delta_\beta)\,, \nonumber\\
&&\CL{e\gamma}{\mu e}^\prime
=\frac{3}{2}(1-\sqrt{3})
{\tilde \beta_e}\left (1-\frac{1+\delta_\alpha}{1+\delta_\beta}
\right )
\simeq \frac{3}{2}(1-\sqrt{3})
{\tilde \beta_e}(\delta_\beta-\delta_\alpha)\,. 
\label{Cdeviation}
 \end{eqnarray}
Therefore, 
 ratios of the Wilson coefficients are expected as:
 \begin{eqnarray}
 \frac{ \CL{e\gamma}{\tau e}^\prime}{\CL{e\gamma}{\mu e}^\prime} \simeq \frac{1}{\sqrt{3}}
 \,, \qquad\qquad
 \frac{ \CL{e\gamma}{\tau e}^\prime}{\CL{e\gamma}{\tau \mu}^\prime} \simeq
 \frac{\tilde \beta_e \delta_\beta}{\tilde \alpha_e \delta_\alpha}\simeq
 \frac{\tilde \beta_e \frac{c}{\tilde \beta_{e(m)}}}{\tilde \alpha_e \frac{c}{\tilde \alpha_{e(m)}}}\simeq  1\,.
 \label{case-2}
 \end{eqnarray}
{It results in} $\mathcal{B}(\tau \to \mu \gamma):
 \mathcal{B}(\tau \to e \gamma):
 \mathcal{B}(\mu \to e \gamma)\sim 1:1:10$ 
 for  the case of  $c_\alpha=c_\beta=c_\gamma=c$.

\subsection{RG evolution contribution of the leptonic dipole operator}
\label{RG}
In the previous subsection, our discussion does not include the {renormalization group (RG)}
contribution.
Let us briefly summarize
 the RG evolution contribution  of the leptonic dipole {operators}  in Ref.\cite{Isidori:2021gqe}  to discuss the RG effect 
 on the numerical results in Eqs.\eqref{case-1} and \eqref{case-2} 
 at the low-energy.
Adopting the SMEFT Warsaw basis \cite{Grzadkowski:2010es}
for the {dimension} $6$ effective operators,
the relevant terms are decomposed as 
\begin{align}
\Delta {\cal  L}_{\rm unbroken}=\Delta {\cal  L}_H +\Delta {\cal  L}_{4f} + {\rm h.c.}\,,
\end{align}	
where
\begin{align}
&\Delta {\cal  L}_H=
-[Y_e]_{pr}(\bar\ell_p e_r) H
+\frac{1}{\Lambda^2} 
\left  [C_{\substack{eH\\pr}}(\bar\ell_p e_r H)(H^\dagger H)
+C_{\substack{eW\\pr}}(\bar\ell_p \sigma^{\mu\nu}e_r)\tau^I H W_{\mu\nu}^I 
+
C_{\substack{eB\\pr}}(\bar\ell_p \sigma^{\mu\nu}e_r) H B_{\mu\nu}
\right ] ,
\nonumber\\
& \Delta {\cal  L}_{4f} =\frac{1}{\Lambda^2}  \left [
C^{(3)}_{\substack{lequ\\prst}}(\bar \ell_p^j \sigma_{\mu\nu} e_r)\epsilon_{jk}(\bar q_s^k \sigma^{\mu\nu} u_t)
+
C^{(1)}_{\substack{lequ\\prst}}(\bar \ell_p^j e_r)\epsilon_{jk}(\bar q_s^k u_t) +C_{\substack{ledq\\prst}}(\bar \ell _p^j e_r)(\bar d_s q_{tj})
\right ]\,.
\label{Lagrangian}
\end{align}	
After the spontaneous breaking of  the Higgs field $H$,
$\Delta {\cal  L}_H$ is rewritten as:
\begin{align}
&\Delta {\cal  L}_{H\,\rm broken}=
-[{\cal Y}_e]_{pr}\frac{v}{\sqrt{2}}(\bar e_{Lp} e_{Rr}) 
+\frac{1}{\Lambda^2} \left[
-[{\cal Y}_{he}]_{pr}\frac{h}{\sqrt{2}}(\bar e_{Lp} e_{Rr}) 
+{\cal C}_{\substack{e\gamma\\pr}}(\bar e_{Lp} \sigma^{\mu\nu}e_{Rr})F_{\mu\nu} \right.
\nonumber\\
& \left . \hskip 2.3 cm +\,{\cal C}_{\substack{eZ\\pr}}\frac{v}{\sqrt{2}}
(\bar e_{Lp}\sigma^{\mu\nu}e_{Rr})
+
{\cal O} (h^2, hF_{\mu\nu},h Z_{\mu\nu}) \right ]\,,
\end{align}	
where $Z_{\mu\nu}$ is the field strength tensor for the $Z$ boson and $h$ is the physical Higgs boson.
The relations between terms in the broken and unbroken {phases} are
\begin{align}
\begin{pmatrix}
{\cal C}_{\substack{e\gamma\\pr}}\\
{\cal C}_{\substack{eZ\\pr}}
\end{pmatrix}=
\begin{pmatrix}
c_\theta & -s\theta\\
-s_\theta & -c\theta\\
\end{pmatrix}
\begin{pmatrix}
{ C}_{\substack{eB\\pr}}\\
{ C}_{\substack{eW\\pr}}
\end{pmatrix}
\,,\qquad
\begin{pmatrix}
[{\cal Y}_e]_{rs}\\
[{\cal Y}_{he}]_{rs}
\end{pmatrix}
=
\begin{pmatrix}
1 & -\frac12\\
1 & \frac23\\
\end{pmatrix}
\begin{pmatrix}
[{ Y}_e]_{rs}\\
v^2 C_{\substack{eH\\rs}}
\end{pmatrix}\,,
\end{align}
{with} 
\begin{align}
c_\theta=\frac{g_2}{\sqrt{g_1^2+g_2^2}}=\frac{e}{g_1}\,,\qquad\quad
s_\theta=\frac{g_1}{\sqrt{g_1^2+g_2^2}}=\frac{e}{g_2}\,,
\end{align}
where $g_1$ and $g_2$ are $U(1)_Y$ and $SU(2)$ gauge couplings.

The solution to {RG equations} of the electromagnetic dipole operators and Yukawa couplings are given  at the one-loop level as follows:
\begin{align}
&{\cal C}_{\substack{e\gamma\\rs}}(\mu_L)
=[1-3\hat L (y_t^2+y_b^2)] \, {\cal C}_{\substack{e\gamma\\rs}}(\mu_H)
-[16\hat Ly_t e]\,{\cal C}^{(3)}_{\substack{\ell equ\\rs33}}(\mu_H)\,,
\\
&[{\cal Y}_e]_{rs}(\mu_L)=
[{Y}_e]_{rs}(\mu_H)-\frac{v^2}{2\Lambda^2}{ C}_{\substack{eH\\rs}}(\mu_H)
+\frac{6v^2}{\Lambda^2}\hat L
[y_t^3 C^{(1)}_{\substack{\ell equ\\rs33}}(\mu_H)
-y_b^3 C_{\substack{\ell edq\\rs33}}(\mu_H)
+\frac34(y_t^2+y_b^2) C_{\substack{eH\\rs}}(\mu_H)]\,,\nonumber
\end{align}
where
\begin{align}
\hat L=\frac{1}{16\pi^2} \log \left ( \frac{\mu_H}{\mu_L}\right )\,,
\end{align}
and $\mu_{H(L)}$ denotes the higher (lower) mass scale.
The coefficient in front of  $C_{ledq}$ {is}  numerically small.   The coefficients $C_{eH}$ controls the $\mu$--$e$ {flavor} violating coupling of the physical Higgs boson, which is tightly constrained 
by other observables~\cite{Blankenburg:2012ex,Harnik:2012pb}  and can be safely ignored in the present analysis.

Finally, approximate evolutions are obtained as follows:
\begin{align}
\label{RGEs}
&{\cal C}_{\substack{e\gamma\\rs}}(\mu_L)
=[1-3\hat L (y_t^2+y_b^2)] \, {\cal C}_{\substack{e\gamma\\rs}}(\mu_H)
-16\hat Ly_t e\,{\cal C}^{(3)}_{\substack{\ell equ\\rs33}}(\mu_H)\,,
\nonumber\\
&[{\cal Y}_e]_{rs}(\mu_L)=
[{Y}_e]_{rs}(\mu_H)
+6\frac{v^2}{\Lambda^2}\hat L
y_t^3 C^{(1)}_{\substack{\ell equ\\rs33}}(\mu_H)
\,.
\end{align}
It is emphasized that  the leptonic flavor structures of 
${\cal C}^{(3)}_{\substack{\ell equ\\rs33}}(\mu_H)$ 
and ${\cal C}^{(1)}_{\substack{\ell equ\\rs33}}(\mu_H)$ 
 are just the same ones as 
${\cal C}_{\substack{e\gamma\\rs}}(\mu_H)$ 
because these Wilson coefficients of the 4-fermion operator 
are  written by a product of 3-point coupling of leptons and that of quarks
 in our Ansatz  Eq.\eqref{eq:Ansatz}.
Therefore, the {RG} contributions do not change
the flavor structure of ${\cal C}_{\substack{e\gamma\\rs}}(\mu_H)$
apart from the overall factor at  low-energy.

On the other hand, 
$[{\cal Y}_e]_{rs}(\mu_L)$ has non-trivial {RG} contribution 
to the flavor structure 
due to $ C^{(1)}_{\substack{\ell equ\\rs33}}(\mu_H)$,
which has   the same flavor structure of 
${\cal C}_{\substack{e\gamma\\rs}}(\mu_H)$. 
If magnitudes of  $ C^{(1)}_{\substack{\ell equ\\rs33}}(\mu_H)$
and  ${\cal C}^{(3)}_{\substack{\ell equ\\rs33}}(\mu_H)$
are comparable, we have  the relation  by using the numerical value of \eqref{Cmumuexp},
\begin{align}
\left |\frac{6\frac{v^2}{\Lambda^2
	}\hat L
	y_t^3 \,C^{(1)}_{\substack{\ell equ\\rs33}}(\mu_H)}{m_\mu/v}\right |\simeq 10^{-3}\times\left |\frac{16\hat Ly_t e\,{\cal C}^{(3)}_{\substack{\ell equ\\rs33}}(\mu_H)}
 {{\cal C}_{\substack{e\gamma\\\mu\mu}}(\mu_L)}\right | 
\,,
\end{align}
where both sides denote  relative contributions of the {RG} 
versus diagonal (2,2) components of 
${\cal C}_{\substack{e\gamma\\rs}}(\mu_L)$ and $[{\cal Y}_e]_{rs}(\mu_L)$,
respectively.
Thus,  the impact  of the term 
 $6v^2\hat L y_t^3 C^{(1)}_{\substack{\ell equ\\rs33}}(\mu_H)$
  on the flavor structure is  minor  in Eq.\eqref{RGEs}
as far as  the {RG} terms are   next-to-leading ones.
Therefore, our numerical result in section \ref{sec:pheno}
is still available even if the {RG} effect is included.

\subsection{EDM of the electron}
The current experimental limit for the electric dipole moment of the electron is given by
ACME collaboration \cite{Andreev:2018ayy}:
\begin{equation}
|d_e/e|~\lesssim~1.1\times10^{-29}\,\text{cm}=5.6\times 10^{-13}\, {\rm TeV^{-1}},
\label{EDMlimit}
\end{equation}
at 90\% confidence level.
Precise measurements of  the electron EDM are rapidly being updated.
The future sensitivity at ACME $\rm I\hskip -0.04cm I\hskip -0.04cm I$ is
 \cite{Kara:2012ay,ACMEIII}:
\begin{equation}
|d_e/e|~\lesssim~0.3\times10^{-30}\,\text{cm}=1.5\times 10^{-14}\,  {\rm TeV^{-1}}.
\label{EDMlimit2}
\end{equation}


The EDM of the electron $d_e$ is defined in the operator:
\begin{align}
\mathcal{O}_{\mathrm{edm}}
= -\frac{i}{2}\,d_e(\mu) \, \overline{e}  \sigma^{\mu\nu} \gamma_5 e F_{\mu\nu}\,,
\label{EDM}
\end{align}
where $d_e=d_e(\mu=m_e)$.
Therefore, the EDM of the electron is extracted from the effective Lagrangian 
\begin{align}
\mathcal{L}_{\rm EDM}=
\frac{1}{\Lambda^2}\CL{e\gamma}{ee}^\prime \mathcal{O}_{\underset{LR}{e\gamma}}
= \frac{1}{\Lambda^2}\CL{e\gamma}{ee}^\prime \frac{v }{\sqrt{2}}  \overline{e}_{L}  \sigma^{\mu\nu}
e_R F_{\mu\nu}\,,
\label{SMEFT-EDM}
\end{align}
which leads to 
\begin{align}
d_e=-\sqrt{2} \, \frac{v}{\Lambda^2}\, {\rm Im}\, [\CL{e\gamma}{ee}^\prime] \,,
\label{SMEFT-EDM}
\end{align}
at tree level, where the small effect of running below the electroweak scale is neglected.
The experimental upper bound in Eq.\eqref{EDMlimit} leads to:
\begin{equation}
 \frac{1}{\Lambda^2}{\mathrm{Im}}\, [\CL{e\gamma}{ee}^\prime]<1.6\times 10^{-12}\, 
 \mathrm {TeV^{-2}}\,,
\label{EDMlimit3}
\end{equation}
{and it}  may be compared with its real part of 
\begin{equation}
 \frac{1}{\Lambda^2}{\mathrm{Re}}\, [\CL{e\gamma}{ee}^\prime]=4.9\times 10^{-8}\, 
 \mathrm {TeV^{-2}}\,,
\label{Real-limit}
\end{equation}
which is derived by  Eqs.\eqref{CG-ee} and \eqref{Cmumuexp}.
What is the origin of the tiny imaginary part of
 $\CL{e\gamma}{ee}^\prime$? 
The coefficient $\CL{e\gamma}{ee}^\prime$ in  Eq.\eqref{C-diagonal}
is rewritten in  a term of small parameter $\delta_\beta$ likewise Eq.\eqref{deviation}
\begin{eqnarray}
\CL{e\gamma}{e e}^\prime=3 \,(1-\sqrt{3})\tilde\beta_e |\epsilon_1^*|
=3 \,(1-\sqrt{3})\tilde\beta_{e(m)}(1+\delta_\beta) |\epsilon_1^*| \,,
\label{Cee-diagonal}
\end{eqnarray}
where $\beta_{e(m)} $ is taken to be real positive by the redefinition of
the right-handed charged lepton field in order to reproduce real positive charged lepton mass.
However, $\delta_\beta$, which is  originated from the unknown mode of
$m$,  is complex in general.  The small parameter
  $\delta_\beta$ could be related to both the $\mu\to e\gamma$
  transition and the electron EDM.
   Eqs.\eqref{Cdeviation} and \eqref{Cee-diagonal} lead to
\begin{eqnarray}
{\rm Im}\, [\CL{e\gamma}{e e}^\prime]
\simeq 3 \,(1-\sqrt{3})\tilde\beta_{e(m)} ({\rm Im}\,\delta_\beta) |\epsilon_1^*| \,,
\qquad 
  \CL{e\gamma}{\mu e}^\prime
  \simeq \frac{3}{2}(1-\sqrt{3})
  {\tilde \beta_{e(m)}}(\delta_\beta-\delta_\alpha)\,. 
  \label{}
\end{eqnarray}
Putting the constraints of experiments in Eqs.\eqref{EDMlimit} and \eqref{Real-limit}, we obtain
\begin{eqnarray}
\frac{{\rm Im}\, [\CL{e\gamma}{e e}^\prime]}
{{\rm Re}\, [\CL{e\gamma}{e e}^\prime]} \simeq 
 ({\rm Im}\,\delta_\beta) < \frac{1.6\times 10^{-12}}{4.9\times 10^{-8}}
=3.3\times 10^{-5}\,. 
\label{delta-beta}
\end{eqnarray}
Suppose $|{\rm Im}\,\delta_\beta|\simeq  |\delta_\beta|$ and
$|\delta_\alpha|\simeq  |\delta_\beta|$ 
(or $|\delta_\alpha|\ll |\delta_\beta|$  ), then, 
this bound is stronger than  $1.4\times 10^{-3}$  from the $\mu\to e \gamma$ in Eq.\eqref{LFV-constraint}.
Indeed,
 the upper bound of the electron EDM forces the branching ratio of $\mu\to e \gamma$ 
 to be
   $\mathcal{B}\!\brackets{\mu^+ \to e^+ \gamma} < 2.3 \times 10^{-16}$.
 
The muon and the tauon EDM may be interesting
 in   high-energy model building \cite{Crivellin:2018qmi}.
   We can also estimate them by using Eq.\eqref{C-diagonal}.
   It is easily found that the predicted value increases at most proportional 
   to its mass. The muon EDM is predicted to be far smaller than the present upper bound \cite{Ema:2021jds}.

Another origin of the electron EDM  is possible
at  one-loop  in the dimension-6 SMEFT. 
The contributions  have been studied
{comprehensively} by discussing 
the  five types of  operators
 $\psi^2HF$, $H^2 F^2$,  $F^3$, $\psi^4$ and $\psi^2 \bar \psi^2$
  in Ref.\cite{Kley:2021yhn}.
 In our scheme of the modular symmetry, 
  the finite contribution comes from the operator
 \cite{Kley:2021yhn}
 \begin{equation}
 \cO_{\underset{abcd}{\ell e}}=(\bar E_L^a\gamma_\mu E_L^b)(\bar E_R^c\gamma_\mu E_R^d)=
 2 (\bar E_L ^aE_R^d)(\bar E_R^c  E_L^b)+E_{LR}^{(2)} \,,
 \end{equation}
 where $E_{LR}^{(2)}$  is an evanescent operator that vanishes 
 in 4 dimensions. 
 Then, we have 
\begin{equation}
\frac{d_e}{e}=-Q_e  \frac{1}{\Lambda^2} \sum_{i = \mu,\tau} \frac{m_i}{8\pi^2} {\mathrm{Im}}\, [\CL{\ell e}{eiie}^\prime]\,,
\end{equation}
where $m_i$ is the muon or tauon mass and 
 $ {\mathrm{Im}}\, [\CL{\ell e}{eiie}^\prime]$ is the Wilson coefficients
 corresponding {to} the operator $ \cO_{\underset{eiie}{\ell e}}$.
Some part of  $ {\mathrm{Im}}\, [\CL{\ell e}{eiie}^\prime]$
is calculable in our scheme.
 After omitting $E_{LR}^{(2)}$,
 the Wilson coefficient  of the operator
 $2 (\bar E_L^e E_R^e)(\bar E_R^i  E_L^i)$ is calculable
  under our Ansatz  Eq.\eqref{eq:Ansatz}
  because 4-point couplings of matter fields are written
  by a product of 3-point couplings of matter fields. 
 The following $A_4$ decomposition
 gives the relevant Wilson coefficients:
  \begin{align}
&2 \left ([\bar E_LY^*(\tau) E_R]_1 \otimes [\bar E_R Y(\tau) E_L]_1+
[\bar E_LY^*(\tau) E_R]_{1'}\otimes [\bar E_R Y(\tau) E_L]_{1''}\right .
\nonumber\\
&+ \left .
 [\bar E_LY^*(\tau) E_R]_{1''}\otimes [\bar E_R Y(\tau) E_L]_{1'}
 +\left \{[\bar E_LY^*(\tau) E_R]_{3_{s},3_a}\otimes [\bar E_R Y(\tau) E_L]_{3_{s},3_a}\right \}_1
\right )\,,
\label{eq:4point}
\end{align}
where the subscripts $1$, $1'$, $1''$, $3_s$ and $3_a$ denote the $A_4$
 representations, respectively.

 The first term is given by  Eqs.\eqref{RLmatrix} and \eqref{LRmatrix}.
 In the mass basis of the charged leptons,
  Eq.\eqref{C-diagonal} gives
 \begin{equation}
\CL{\ell e}{e\mu\mu e}^\prime= 
\CL{e\gamma}{e e}^\prime \times \CL{e\gamma}{\mu\mu}^\prime
=\frac29 (1-\sqrt{3})^2\tilde\alpha_e \tilde\beta_e
|\epsilon_1^*|\,,
\qquad
\CL{\ell e}{e\tau\tau e}^\prime= 
\CL{e\gamma}{e e}^\prime \times \CL{e\gamma}{\tau\tau}^\prime
=3\sqrt{3}(1-\sqrt{3})^2\tilde\gamma_e \tilde\beta_e
|\epsilon_1^*|\,.
 \end{equation}
 In order to calculate the magnitude of these coefficients,
 which is the NP contribution, we estimate  $\tilde\alpha_e$
 up to the overall strength by
combining Eqs.\eqref{C-diagonal} and \eqref{Cmumuexp} as: 
  \begin{equation}
  \frac{1}{\Lambda^2}{\rm Re}\, [\CL{e\gamma}{\mu\mu}^\prime]
  =\frac{1}{\Lambda^2}\,\frac32 (\sqrt{3}-1){\rm Re}\,\tilde\alpha_e =1.0\times 10^{-5}\,  {\rm TeV}^{-2}\,\  \Rightarrow \  
   {\rm Re}\,\tilde\alpha_e =\frac{2}{3 (\sqrt{3}-1)}\times 10^{-5}
  \left (\frac{\Lambda}{1\, {\rm TeV}}\right )^2 \,.
 \end{equation}
The magnitudes of  $\tilde\beta_e$, $\tilde\gamma_e$ 
and $|\epsilon_1^*|$ are obtained by the relations  in Eq.\eqref{Lepton-model} since $\tilde\alpha_e$,
 $\tilde\beta_e$, $\tilde\gamma_e$  are almost real.
Then, we obtain  the tauon contribution as:
\begin{equation}
\CL{\ell e}{e\tau\tau e}^\prime=8.6 \times 10^{-12} 
\left (\frac{\Lambda}{1\, {\rm TeV}}\right )^4 \ e^{i \phi}\,,
\end{equation}
 where $\phi$ is the phase
 originated from small imaginary parts of  $\tilde\alpha_e$,
 $\tilde\beta_e$, $\tilde\gamma_e$. 
 On the other hand, the muon contribution
 $\CL{\ell e}{e\mu\mu e}^\prime$ is much smaller than that.
 Then, we have  
\begin{equation}
\frac{d_e}{e}\simeq -Q_e  \frac{1}{\Lambda^2}  \frac{m_\tau}{8\pi^2} {\mathrm{Im}}\, [\CL{\ell e}{e\tau\tau e}^\prime]
\simeq 2\times 10^{-16}
\left (\frac{\Lambda}{1\, {\rm TeV}}\right )^2\sin\phi 
\ \  [{\rm TeV^{-1}}]\,,
\label{edm-mtau}
\end{equation}
where $\sin\phi$ is at most $10^{-5}$ as discussed in Eq.\eqref{delta-beta}. 
{Here, the cutoff scale $\Lambda$ is upper bounded by ${\cal O}(100\, {\rm TeV})$; 
otherwise the magnitude of $\tilde \gamma_e$ exceeds ${\cal O}(1)$ due to 
the relation (\ref{Cmumuexp}).} 

 However, the second and {third} terms {in Eq. (\ref{eq:4point})} could include the phase of ${\cal O}(1)$
  because they do not relate with the mass matrix of
 charged leptons. In the basis of real positive diagonal masses,
 there are complex in general.
 Those terms are obtained by the cyclic permutation $1\to 1'$, $1'\to 1''$ and $1''\to 1$ in Eq.\eqref{RLmatrix} and taking its dagger,
 in which 
 ($\tilde\alpha_e$, $\tilde\beta_e$, $\tilde\gamma_e$) are replaced
 with complex new parameters ($\tilde\alpha'_e$, $\tilde\beta'_e$, $\tilde\gamma'_e$) and ($\tilde\alpha''_e$, $\tilde\beta''_e$, $\tilde\gamma''_e$), respectively.
 Although new parameters are unknown ones, we can assume reasonably
  them to be comparable  to
 ($\tilde\alpha_e$, $\tilde\beta_e$, $\tilde\gamma_e$).
 Therefore, the electron EDM is estimated roughly
  by  Eq.\eqref{edm-mtau} by putting $\sin\phi\sim 1$.
  Then, $d_e/e$ is expected to be
   $2\times 10^{-14}\,(5\times 10^{-13})\, {\rm TeV^{-1}}$ for $\Lambda=10\,(50)\, {\rm TeV}$.
  These are  consistent with  the present upper-bound 
   $d_e/e< 5.6\times 10^{-13} \,{\rm TeV}^{-1}$.
 {It is noted that the last term of $3\otimes 3$ 
 	{in Eq.(\ref{eq:4point})} is unable to be
 estimated because this term has many unknown parameters.}


\section{Summary}
\label{sec:summary}
We have studied the  LFV decays,  the  anomalous magnetic moments $(g-2)_{\mu,\, e}$ and  the EDM of the electron 
in  the SMEFT with the $\Gamma_N$ modular flavor symmetry. 
We employ the relation Eq.(\ref{eq:Ansatz}) as Ansatz in the SMEFT.
Through this Ansatz, higher-dimensional operators are related with 3-point couplings.
 We take the level 3 finite modular group, 
$\Gamma_3$ for the flavor symmetry, and 
discuss the dipole  operators  at nearby fixed point $\tau=i$,
where observed  lepton masses and mixing angles are well reproduced.


Suppose the anomaly of the anomalous magnetic moment of the muon,
 $\Delta a_\mu$ to be  evidence of NP, we have  related it with 
 the electron $\Delta a_e$, the LFV decays and the {electron} EDM.
 It is found that the NP contribution to $a_{e}$
 and $a_{\mu}$ {is} proportional to the lepton {masses squared} likewise  the naive scaling 
 $\Delta a_\ell \propto m^2_\ell$.
 {The} predicted value of the anomaly of $(g-2)_{e}$ is small of one order compared with the observed one at present. 
 The {precise} observation of the fine structure constant
 will provide the test of our framework.
 

We have also predicted  the correlations among
the LFV processes: $\mu\to e\gamma$, $\tau\to \mu\gamma$
and $\tau\to e\gamma$. 
They depend on the additional unknown mode of $m$ in Eq.(\ref{eq:Ansatz}).
If the unknown mode is Higgs-like,
 we have predicted 
$\mathcal{B}(\tau \to \mu \gamma):
\mathcal{B}(\tau \to e \gamma):
\mathcal{B}(\mu \to e \gamma)\sim 10^4:1:10$.
On the other hand,
if  the   unknown mode   is the flavor blind,
$\mathcal{B}(\tau \to \mu \gamma):
\mathcal{B}(\tau \to e \gamma):
\mathcal{B}(\mu \to e \gamma)\sim 1:1:10$
is predicted.
The experimental test will be available  in the future.
It is also remarked that our numerical result of the LFV processes
is  acceptable even if the {RG} effect is included.


In order to realize the electron EDM,  we need the tiny imaginary part of
the Wilson coefficient 
$\CL{e\gamma}{ee}^\prime$. 
In the basis of real positive charged lepton masses, 
 the imaginary part originates from the unknown mode of
$m$, which is complex in general.
  The parameters of the unknown mode
 could be related to both the $\mu\to e\gamma$
transition and the electron EDM.
It is found that 
the upper bound of the electron EDM forces the  $\mu\to e \gamma$ 
decay to be
$\mathcal{B}\!\brackets{\mu^+ \to e^+ \gamma} < 2.3 \times 10^{-16}$.

Another origin of the electron EDM  is possible
at  one-loop level in the dimension-6 SMEFT. 
The finite contribution comes from
the Wilson coefficient  of the operator 
$2 (\bar E_L^e E_R^e)(\bar E_R^i  E_L^i)$.
Then, we have obtained 
${d_e}/{e}\simeq
 2\times 10^{-16} \Lambda^2 \sin\phi $,
where  $\Lambda$ denotes a certain mass scale,
and $\phi$ is an unknown phase of ${\cal O}(1)$ in the Wilson coefficient.
Then, $d_e/e$ is expected to be
$2\times 10^{-14}\,(5\times 10^{-13})\, {\rm TeV^{-1}}$ for $\Lambda=10\,(50)\, {\rm TeV}$.
These are  consistent with  the present upper bound 
$d_e/e< 5.6\times 10^{-13} \,{\rm TeV}^{-1}$.

Thus, our Ansatz in the SMEFT with the modular symmetry of flavors
is {powerful} to study the leptonic phenomena of flavors
{comprehensively}.

\vspace{0.5 cm}
\noindent
{\large\bf Acknowledgement}\\
This work was supported by  JSPS KAKENHI Grant Numbers JP20K14477 (HO), and JP21K13923 (KY).

\appendix
\section*{Appendix}

\section{Experimental constraints on the dipole operators}
\label{sect:exp}
We summarize {briefly}  the experimental constraints on the dipole operators given by Ref.\cite{Isidori:2021gqe}.
Below the scale of electroweak symmetry breaking,
the leptonic dipole operators {are} given as:
\begin{align}
\mathcal{O}_{\underset{rs}{e\gamma}}
= \frac{v }{\sqrt{2}}  \overline{e}_{L_r}  \sigma^{\mu\nu} e_{R_s} F_{\mu\nu}\,,
\label{eq:dipoledef}
\end{align}
where $\{r,s\}$ are  {flavor} indices $e,\mu,\tau$ and $F_{\mu\nu}$ is the electromagnetic field strength tensor. 
The corresponding Wilson coefficient is denoted by $\CL{e\gamma}{rs}^\prime$
in the mass basis of leptons.

The combined result from the E989 experiment at FNAL~\cite{Muong-2:2021ojo} and the E821 experiment at BNL~\cite{Muong-2:2006rrc} on
the   $a_\mu=(g-2)_\mu/2$, 
together with the SM prediction in~\cite{Aoyama:2020ynm}, {implies} 
\begin{align}
\Delta a_\mu &= a_\mu^\mathrm{Exp} - a_\mu^\mathrm{SM} = \brackets{251 \pm 59} \times 10^{-11}~.
\label{muon-data}
\end{align}
The tree-level expression for  $\Delta a_\mu$ in terms of the Wilson coefficient of the dipole operator is 
\begin{align}
\Delta a_{\mu} &= \frac{4 m_{\mu}}{e}   \frac{v}{\sqrt 2} \,\frac{1}{\Lambda^2}\text{Re} \, [\CL{e\gamma}{\mu\mu}^\prime] \,,
\label{eq:magnetic-moment}
\end{align}
where
 $v\approx 246$~GeV and $\Lambda$ is a certain mass scale of NP.
Here the Wilson coefficient is understood to be evaluated at the weak scale 
(we neglect the small effect of running below the weak scale),
and the  prime of the Wilson coefficient  indicates the {flavor} basis corresponding to the mass-eigenstate basis of charged leptons \footnote{The one-loop relation can be found in~\cite{Buttazzo:2020ibd	}.}.
Inputting the experimental results  leads to 
\begin{align}
\frac{1}{\Lambda^2}\text{Re}\  [\CL{e\gamma}{\mu\mu}^\prime] \approx  1.0 \times 10^{-5} \, \mathrm{TeV}^{-2} \, .
\label{Cmumuexp}
\end{align}

The tree-level expression of a  radiative LFV rate in terms of the Wilson coefficients  is 
\begin{align}
\mathcal{B}(\ell_r \to \ell_s \gamma) = \frac{m_{\ell_r}^3 v^2}{8 \pi \Gamma_{\ell_r}} \frac{1}{\Lambda^4}\left(|\CL{e\gamma}{rs}'|^2 + |\CL{e\gamma}{sr}'|^2\right) \, .
\label{eq:Branching-ratio_lepton-decay}
\end{align}
Using this expression, the experimental bound $\mathcal{B}\!\brackets{\mu^+ \to e^+ \gamma} < 4.2 \times 10^{-13}$~(90\%~C.L.)
obtained by the MEG experiment~\cite{TheMEG:2016wtm} can be translated into the upper bound
\begin{align}
\frac{1}{\Lambda^2}|\CL{e\gamma}{e\mu(\mu e)}^\prime| <  2.1 \times 10^{-10} \, \mathrm{TeV}^{-2} \, .
\label{eq:bound_C_egamma_12}
\end{align}
Taking into account Eqs.~(\ref{Cmumuexp}) and \eqref{eq:bound_C_egamma_12}, we have the ratio:
\begin{align}
\left |\frac{ \CL{e\gamma}{e\mu(\mu e)}^\prime  }{    \CL{e\gamma}{\mu\mu}^\prime    }\right | <  2.1\times 10^{-5}\,.
\label{eq:bound12}
\end{align}


\section{$A_4$ modular symmetry}
\label{sec:mod}
\subsection{Modular flavor symmetry}
We briefly review the models with $A_4$ modular symmetry.
The modular group $\bar\Gamma$ is the group of linear fractional transformations
$\gamma$ acting on the modulus  $\tau$, 
belonging to the upper-half complex plane as:
\begin{equation}\label{eq:tau-SL2Z}
\tau \longrightarrow \gamma\tau= \frac{a\tau + b}{c \tau + d}\ ,~~
{\rm where}~~ a,b,c,d \in \mathbb{Z}~~ {\rm and }~~ ad-bc=1, 
~~ {\rm Im} [\tau]>0 ~ ,
\end{equation}
which is isomorphic to  $PSL(2,\mathbb{Z})=SL(2,\mathbb{Z})/\{\rm I,-I\}$ transformation.
This modular transformation is generated by $S$ and $T$, 
\begin{eqnarray}
S:\tau \longrightarrow -\frac{1}{\tau}\ , \qquad\qquad
T:\tau \longrightarrow \tau + 1\ ,
\label{symmetry}
\end{eqnarray}
which satisfy the following algebraic relations, 
\begin{equation}
S^2 ={\rm I}\ , \qquad (ST)^3 ={\rm I}\ .
\end{equation}

We introduce the series of groups $\Gamma(N)$, called principal congruence subgroups, where  $N$ is the level $1,2,3,\dots$.
These groups are defined by
\begin{align}
\begin{aligned}
\Gamma(N)= \left \{ 
\begin{pmatrix}
a & b  \\
c & d  
\end{pmatrix} \in SL(2,\mathbb{Z})~ ,
~~
\begin{pmatrix}
a & b  \\
c & d  
\end{pmatrix} =
\begin{pmatrix}
1 & 0  \\
0 & 1  
\end{pmatrix} ~~({\rm mod} N) \right \}
\end{aligned} .
\end{align}
For $N=2$, we define $\bar\Gamma(2)\equiv \Gamma(2)/\{\rm I,-I\}$.
Since the element $\rm -I$ does not belong to $\Gamma(N)$
for $N>2$, we have $\bar\Gamma(N)= \Gamma(N)$.
The quotient groups defined as
$\Gamma_N\equiv \bar \Gamma/\bar \Gamma(N)$
are  finite modular groups.
In these finite groups $\Gamma_N$, $T^N={\rm I}$  is imposed.
The  groups $\Gamma_N$ with $N=2,3,4,5$ are isomorphic to
$S_3$, $A_4$, $S_4$ and $A_5$, respectively \cite{deAdelhartToorop:2011re}.

Modular forms $f_i(\tau)$ of weight $k$ are the holomorphic functions of $\tau$ and transform as
\begin{equation}
f_i(\tau) \longrightarrow (c\tau +d)^k \rho(\gamma)_{ij}f_j( \tau)\, ,
\quad \gamma\in \bar \Gamma\, ,
\label{modularforms}
\end{equation}
under the modular symmetry, where
$\rho(\gamma)_{ij}$ is a unitary matrix under $\Gamma_N$.

Under the modular transformation of Eq.\,(\ref{eq:tau-SL2Z}), chiral superfields $\psi_i$ ($i$ denotes flavors) with weight $-k$
transform as \cite{Ferrara:1989bc}
\begin{equation}
\psi_i\longrightarrow (c\tau +d)^{-k}\rho(\gamma)_{ij}\psi_j\, .
\label{chiralfields}
\end{equation}

We study global SUSY models.
The superpotential which is built from matter fields and modular forms
is assumed to be modular invariant, i.e., to have 
a vanishing modular weight. For given modular forms, 
this can be achieved by assigning appropriate
weights to the matter superfields.

The kinetic terms  are  derived from a K\"ahler potential.
The K\"ahler potential of chiral matter fields $\psi_i$ with the modular weight $-k$ is given simply  by 
\begin{equation}\label{eq:Kahler}
\frac{1}{[i(\bar\tau - \tau)]^{k}} \sum_i|\psi_i|^2,
\end{equation}
where the superfield and its scalar component are denoted by the same letter, and $\bar\tau =\tau^*$ after taking VEV of $\tau$.
The canonical form of the kinetic terms  is obtained by 
changing the normalization of parameters \cite{Kobayashi:2018scp}.
The general K\"ahler potential consistent with the modular symmetry possibly contains additional terms \cite{Chen:2019ewa}. However, we consider only the simplest form of
the K\"ahler potential, {as naturally realized in the effective action of higher-dimensional theory such as the string theory.} 

For $\Gamma_3\simeq A_4$, the dimension of the linear space 
${\cal M}_k(\Gamma{(3)})$ 
of modular forms of weight $k$ is $k+1$ \cite{Gunning:1962,Schoeneberg:1974,Koblitz:1984}, i.e., there are three linearly 
independent modular forms of the lowest non-trivial weight $2$,
which form a triplet of the $A_4$ group.
These modular forms have been explicitly given \cite{Feruglio:2017spp} in the symmetric base of the $A_4$ generators $S$ and $T$ for the triplet representation
as shown in the next subsection.  


\subsection{Modular forms}
\label{modular-form}
The holomorphic and anti-holomorphic modular forms {of} weight 2 
compose the $A_4$ triplet as: 
\begin{align}
&\begin{aligned}
Y(\tau)=
\begin{pmatrix}
Y_1(\tau)  \\
Y_2(\tau) \\
Y_3(\tau)
\end{pmatrix}\,,\qquad 
\overline {Y(\tau)}\equiv Y^{*}(\tau)=
\begin{pmatrix}
Y_1^*(\tau)  \\
Y_3^*(\tau) \\
Y_2^*(\tau)
\end{pmatrix}\,.
\end{aligned}
\end{align}
In the representation of 
the generators $S$ and $T$  for $A_4$ triplet:
\begin{align}
\begin{aligned}
S=\frac{1}{3}
\begin{pmatrix}
-1 & 2 & 2 \\
2 &-1 & 2 \\
2 & 2 &-1
\end{pmatrix},
\end{aligned}
\qquad\quad 
\begin{aligned}
T=
\begin{pmatrix}
1 & 0& 0 \\
0 &\omega& 0 \\
0 & 0 & \omega^2
\end{pmatrix}, 
\end{aligned}
\label{ST}
\end{align}
where $\omega=e^{i\frac{2}{3}\pi}$,
modular forms  are given explicitly
 \cite{Feruglio:2017spp}:
\begin{eqnarray} 
\label{eq:Y-A4}
Y_1(\tau) &=& \frac{i}{2\pi}\left( \frac{\eta'(\tau/3)}{\eta(\tau/3)}  +\frac{\eta'((\tau +1)/3)}{\eta((\tau+1)/3)}  
+\frac{\eta'((\tau +2)/3)}{\eta((\tau+2)/3)} - \frac{27\eta'(3\tau)}{\eta(3\tau)}  \right), \nonumber \\
Y_2(\tau) &=& \frac{-i}{\pi}\left( \frac{\eta'(\tau/3)}{\eta(\tau/3)}  +\omega^2\frac{\eta'((\tau +1)/3)}{\eta((\tau+1)/3)}  
+\omega \frac{\eta'((\tau +2)/3)}{\eta((\tau+2)/3)}  \right) , \label{Yi} \\ 
Y_3(\tau) &=& \frac{-i}{\pi}\left( \frac{\eta'(\tau/3)}{\eta(\tau/3)}  +\omega\frac{\eta'((\tau +1)/3)}{\eta((\tau+1)/3)}  
+\omega^2 \frac{\eta'((\tau +2)/3)}{\eta((\tau+2)/3)}  \right)\, ,
\nonumber
\end{eqnarray}
where $\eta(\tau)$ is the Dedekind eta function, 
\begin{align}
\eta(\tau) = q^{1/24}\prod^\infty_{n=1} (1-q^n), \qquad q={\rm exp}~(2 \pi i \tau).
\end{align}
Those are also  expressed in the expansions of  $q=\exp (2i\pi\tau)$:
\begin{align}
\begin{pmatrix}Y_1(\tau)\\Y_2(\tau)\\Y_3(\tau)\end{pmatrix}=
\begin{pmatrix}
1+12q+36q^2+12q^3+\dots \\
-6q^{1/3}(1+7q+8q^2+\dots) \\
-18q^{2/3}(1+2q+5q^2+\dots)
\end{pmatrix}\,.
\label{Y(2)}
\end{align}

\section{A model of  mass matrices in  $A_4$ modular symmetry}
\label{massmatrixmodel}
We present a  model for leptons \cite{Okada:2020brs}, where
the neutrino mass matrix is given 
in terms of weight 4  modular forms by using  Weinberg operator.
The prediction of this model is consistent with the  
NuFIT 5.0 data \cite{Esteban:2020cvm}.
The  assignments of representations and modular weights to the lepton fields
are presented in Table \ref{tb:lepton}. 
\begin{table}[H]
	\centering
	\begin{tabular}{|c||c|c|c|cc|} \hline
		\rule[14pt]{0pt}{1pt}
		&$L_L$&$(e^c_R,\mu^c_R,\tau^c_R)$&$H_d,\, H_u$
		&$Y_{\bf r}^{\rm (2)}$ & $Y_{\bf r}^{\rm (4)}$ \\  \hline\hline 
		\rule[14pt]{0pt}{1pt}
		$SU(2)$&$\bf 2$&$\bf 1$&$\bf 2$ & \multicolumn{2}{c|}{$\bf 1$}\\
		\rule[14pt]{0pt}{1pt}
		$A_4$&$\bf 3$& \bf (1,\ 1$''$,\ 1$'$)&$\bf 1$ & $\bf 3$ &$\bf \{3, 1, 1'\} $ \\
		\rule[14pt]{0pt}{1pt}
		$k$& $2$ &$(0,\ 0,\ 0)$ &0 & $2$ & $4$ \\ \hline
	\end{tabular}	
	\caption{ Representations and  weights
		$k$ for MSSM fields and  modular forms of weight $2$ and $4$.
	}
	\label{tb:lepton}
\end{table}

The charged lepton mass matrix and neutrino ones are given as:
\begin{align}
&M_E=v_d
\begin{pmatrix}
\alpha_{e(m)} & 0 & 0 \\
0 &\beta_{e(m)} & 0\\
0 & 0 &\gamma_{e(m)}
\end{pmatrix} 
\begin{pmatrix}
Y_1 & Y_3 & Y_2 \\
Y_2 & Y_1 &  Y_3 \\
Y_3&  Y_2&  Y_1
\end{pmatrix}_{RL}, \nonumber\\
&M_\nu=\frac{v_u^2}{\Lambda} \left [
\begin{pmatrix}
2Y_1^{(4)} & -Y_3^{(4)} & -Y_2^{(4)}\\
-Y_3^{(4)} & 2Y_2^{(4)} & -Y_1^{(4)} \\
-Y_2^{(4)} & -Y_1^{(4)} & 2Y_3^{(4)}
\end{pmatrix}
+g^{\nu}_1 Y_{\bf 1}^{\rm (4)} 
\begin{pmatrix}
1 & 0 &0\\ 0 & 0 & 1 \\ 0 & 1 & 0
\end{pmatrix}
+g^{\nu}_2 Y_{\bf 1'}^{\rm (4)}
\begin{pmatrix}
0 & 0 &1\\ 0 & 1 & 0 \\ 1 & 0 & 0
\end{pmatrix}
\right ] \, ,
\label{ME222}
\end{align}
respectively,
where  $\alpha_{e(m)}$, $\beta_{e(m)}$ and $\gamma_{e(m)}$ are real 
and  $g_1^\nu$,  $g_2^\nu$ are supposed to be also real.
The parameters $v_d$ and $v_u$ denote the VEV of Higgs fields $H_d$ and $H_u$, respectively.


The modular forms of higher weights are obtained by tensor products of 
${ Y^{\rm (2)}_{\bf 3}}(\tau)$.
The modular forms of weight $k=4$ have the dimension $d_4=5$.
They decompose to ${\bf 3}$, ${\bf 1}$, and ${\bf 1}'$ and are written explicitly by 
\begin{align}
&{ Y^{\rm (4)}_{\bf 3}}(\tau)=
\begin{pmatrix}
Y_1^2-Y_2Y_3  \\
Y_3^2-Y_1Y_2 \\
Y_2^2-Y_1Y_3
\end{pmatrix}\,,
\nonumber\\
&Y_{\bf 1}^{(4)}=Y_1^2+2Y_2Y_3\,, \qquad\qquad 
Y_{\bf 1'}^{(4)}=Y_3^2+2Y_1Y_2\,. 
\label{Y4}
\end{align}

\section{Charged lepton mass matrix at nearby $\tau=i$}
\label{massmatrix-i}

Residual symmetries arise whenever the VEV of the modulus $\tau$ breaks
the modular group $\overline{\Gamma}$ only partially.
There are only 2 inequivalent finite points in the fundamental domain
of $\overline{\Gamma}$,
namely,  $\tau = i$ and 
$ \tau =\omega=-1/2+ i \sqrt{3}/2$.
The first point is  invariant under the $S$ transformation
$\tau=-1/\tau$.  In the case of $A_4$ symmetry, the subgroup $\mathbb{Z}_2^{S}=\{ I, S \}$ is preserved at $ \tau = i$.

If  a residual symmetry of $S$  in $A_4$ 
is preserved in  mass matrices of leptons,
we have commutation relations 
between the mass matrices and the generator $S$ as:
\begin{align}
[M_{E}^\dagger M_{E},\, { S}]=0 \, . 
\label{commutator}
\end{align}
Therefore, $M_{E}^\dagger M_{E}$ could be diagonal
in the diagonal basis of the generator $S$.

Let us consider the fixed point $\tau=i$, where
holomorphic and  anti-holomorphic modular forms of weight $2$ are given  as:
\begin{align}
&\begin{aligned}
{ Y}(\tau_q=i)=Y_1(i)
\begin{pmatrix}
1  \\ 1-\sqrt{3} \\ -2+\sqrt{3}
\end{pmatrix}\,,\qquad 
\ Y^{*}(\tau_q=i)=Y_1(i)
\begin{pmatrix}
1  \\ -2+\sqrt{3} \\ 1-\sqrt{3}
\end{pmatrix}\,,
\end{aligned}\nonumber\\
&\begin{aligned}
{ Y}(\tau_e=i)=Y_1(i)
\begin{pmatrix}
1  \\ 1-\sqrt{3} \\ -2+\sqrt{3}
\end{pmatrix}\,,\qquad 
\ Y(\tau_e=i)=Y_1(i)
\begin{pmatrix}
1  \\ -2+\sqrt{3} \\ 1-\sqrt{3}
\end{pmatrix}\,,
\end{aligned}
\label{modularS}
\end{align}
in the basis of Eq.\,\eqref{ST}.

The charged lepton mass matrix of Eq.\,(\ref{ME222}) {is} given in terms of weight $2$ modular forms.
Perform  the unitary transformation
of $E_L\to U_S E_L$, where   
\begin{align}
U_S =
\frac{1}{2\sqrt{3}}
\begin{pmatrix}
2  & 2 & 2\\
\sqrt{3}+1  &  -2 & \sqrt{3}-1\\
\sqrt{3}-1 & -2  & \sqrt{3}+1
\end{pmatrix}\,.
\label{U-S0}
\end{align}
It is  noticed that $U_S S U_S^\dagger$ is diagonal.
Then, the charged lepton mass matrix at $\tau=i$
in Eq.\,(\ref{ME222}) is simply given as:
\begin{align} 
&M_{E}\,=\frac12 v_d
\begin{pmatrix}
0& 3(\sqrt{3}-1) \tilde\alpha_{e(m)} &-(3-\sqrt{3}) \tilde\alpha_{e(m)}\\
0 & -3(\sqrt{3}-1) \tilde\beta_{e(m)} & -(3-\sqrt{3}) \tilde\beta_{e(m)}\\
0 &0& 2(3-\sqrt{3})\tilde\gamma_{e(m)}
\end{pmatrix}_{RL}\,, \nonumber\\
&M_{E}^\dagger M_{E}=\frac12 v_d^2
\begin{pmatrix}
0  & 0 & 0\\
0  &  9(2-\sqrt{3}) (\tilde\alpha_{e(m)}^2+\tilde\beta_{e(m)}^2) &
3(3-2\sqrt{3}) (\tilde\alpha_{e(m)}^2-\tilde\beta_{e(m)}^2)\\
0 &  3(3-2\sqrt{3}) (\tilde\alpha_{e(m)}^2-\tilde\beta_{e(m)}^2) & 
3(2-\sqrt{3}) (\tilde\alpha_{e(m)}^2+\tilde\beta_{e(m)}^2+4\tilde\gamma_{e(m)}^2)
\end{pmatrix}_{LL}\,,
\label{charged-matrix-S}
\end{align}
where $\tilde \alpha_{e(m)}=(6-3\sqrt{3})  Y_1(i)\alpha_{e(m)}$,
$\tilde \beta_{e(m)}= (6-3\sqrt{3})  Y_1(i) \beta_{e(m)}$ and 
$\tilde \gamma_{e(m)} =(6-3\sqrt{3}) Y_1(i)  \gamma_{e(m)}$,
and the magnitede of 
$\tilde \gamma_{e(m)}$ is supposed to be  much larger than  $\tilde \alpha_{e(m)}$
and  $\tilde \beta_{e(m)}$.  
Since two eigenvalues of $S$ are degenerate
such as $(1,-1,-1)$, there is still a freedom of the  $2$--$3$ family rotation.
Therefore, $M_{E}^\dagger M_{E}$ could be diagonal
after the small $2$--$3$ family rotation of
${\cal O}(\tilde \alpha_{e(m)}^2\tilde/ \gamma_{e(m)}^2,\,\tilde \beta_{e(m)}^2/\tilde \gamma_{e(m)}^2)$.


Since the lepton mass matrices cannot reproduce
the observed  PMNS matrices at fixed points as discussed in Ref.\cite{Okada:2020ukr}.
Therefore, the deviations from the fixed points are required
to realize observed masses and mixing angles.
Let us consider the small deviation from $\tau=i$.
By using the approximate modular forms of weight 2
at  $\tau=i+\epsilon$ in Eq.\eqref{epS12}, we present
$M_E^\dagger M_E$ 
including of  order $\epsilon_1\,(\simeq 2.05 \,i\,\epsilon)$,
which gives the mixing angles in the left-handed sector.
Performing the transformation of  $E_L\to   U_{S} E_L$, 
	where $U_S$ is in  Eq.\,\eqref{U-S0},
	we have 
\scriptsize
\begin{align} 
M_{E}^\dagger M_{E}\simeq v_d^2
\begin{pmatrix}
2(2-\sqrt{3}) (\tilde\gamma_{e(m)}^2+\tilde\alpha_{e(m)}^2+\tilde\beta_{e(m)}^2)|\epsilon_1|^2
& 3(\sqrt{3}-2) (\tilde\alpha_{e(m)}^2-\tilde\beta_{e(m)}^2)\epsilon_1^* 
& 
(3-2\sqrt{3}) (2\tilde\gamma_{e(m)}^2-\tilde\alpha_{e(m)}^2-\tilde\beta_{e(m)}^2)\epsilon_1^* \\
3(\sqrt{3}-2) (\tilde\alpha_{e(m)}^2-\tilde\beta_{e(m)}^2)\epsilon_1   & 
\frac92 (2-\sqrt{3}) (\tilde\alpha_{e(m)}^2+\tilde\beta_{e(m)}^2)  &
\frac32 (3-2\sqrt{3}) (\tilde\alpha_{e(m)}^2-\tilde\beta_{e(m)}^2) \\
(3-2\sqrt{3}) (2\tilde\gamma_{e(m)}^2-\tilde\alpha_{e(m)}^2-\tilde\beta_{e(m)}^2)\epsilon_1 & 
\frac32 (3-2\sqrt{3}) (\tilde\alpha_{e(m)}^2-\tilde\beta_{e(m)}^2) & 
\frac32(2-\sqrt{3}) (4\tilde\gamma_{e(m)}^2+\tilde\alpha_{e(m)}^2+\tilde\beta_{e(m)}^2)
\end{pmatrix}
\nonumber\\ \nonumber\\
	=	v_d^2 P_e^*
	\begin{pmatrix}
	2(2-\sqrt{3}) (\tilde\gamma_{e(m)}^2+\tilde\alpha_{e(m)}^2+\tilde\beta_{e(m)}^2)|\epsilon_1|^2
	& 3(\sqrt{3}-2) (\tilde\alpha_{e(m)}^2-\tilde\beta_{e(m)}^2)|\epsilon_1^*|
	& 
	(3-2\sqrt{3}) (2\tilde\gamma_{e(m)}^2-\tilde\alpha_{e(m)}^2-\tilde\beta_{e(m)}^2)
	|\epsilon_1^*| \\
	3(\sqrt{3}-2) (\tilde\alpha_{e(m)}^2-\tilde\beta_{e(m)}^2)|\epsilon_1|   & 
	\frac92 (2-\sqrt{3}) (\tilde\alpha_{e(m)}^2+\tilde\beta_{e(m)}^2)  &
	\frac32 (3-2\sqrt{3}) (\tilde\alpha_{e(m)}^2-\tilde\beta_{e(m)}^2) \\
	(3-2\sqrt{3}) (2\tilde\gamma_{e(m)}^2-\tilde\alpha_{e(m)}^2-\tilde\beta_{e(m)}^2)
	|\epsilon_1| & 
	\frac32 (3-2\sqrt{3}) (\tilde\alpha_{e(m)}^2-\tilde\beta_{e(m)}^2) & 
	\frac32(2-\sqrt{3}) (4\tilde\gamma_{e(m)}^2+\tilde\alpha_{e(m)}^2+\tilde\beta_{e(m)}^2)
	\end{pmatrix} P_e
\,,
\label{Emassmatrix-m}
\end{align}
\normalsize
where   $\gamma_{e(m)}^2\gg \alpha_{e(m)}^2\gg\beta_{e(m)}^2$ is taken 
following from the numerical result in Ref.\cite{Okada:2020brs},
and  $\epsilon_2=2\epsilon_1$ in Eq.\,(\ref{epS12}) is put.
	The phase matrix $P_e$ is given as
	\begin{align}
	P_e=
	\begin{pmatrix}
	e^{i\eta_e} & 0 & 0\\
	0& 1 & 0 \\
	0 & 0 & 1
	\end{pmatrix}
	\,,\qquad\qquad \eta_e= {\rm arg}\,[\epsilon_1] \,.
	\end{align}
The matrix of Eq.\eqref{Emassmatrix-m} is a rank one  at the limit of  $\alpha_{e(m)}^2=\beta_{e(m)}^2=0$. 
Putting small relevant  values of  $\alpha_{e(m)}^2/\gamma_{e(m)}^2$, $\beta_{e(m)}^2/\gamma_{e(m)}^2$
and $|\epsilon_1|$, observed charged lepton masses could be obtained.

	The mixing matrix $U_{Lme}$ 
	to diagonarize $M_{E}^\dagger M_{E}$
	such as $U_{Lme}^\dagger M_{E}^\dagger M_{E}U_{Lme}={\rm diag}(m_e^2, m_\mu^2, m_\tau^2)$ is given as: 
	\begin{align}
	U_{Lme}\simeq  P_e^*
	\begin{pmatrix}
	1 & s_{L12}^{e}  &s_{L13}^{e}\\
	-s_{L12}^{e} & 1 & s_{L23}^{e}\\
	s_{L12}^{e} s_{L23}^{e} -s_{L13}^{e}   & -s_{L23}^{e}  & 1
	\end{pmatrix}\,,
	\end{align}
	where 
	\begin{align}
	s^e_{L12}\simeq -|\epsilon_1^*|\,,\qquad\quad\ \quad  
	s^e_{L23}\simeq -\frac{\sqrt{3}}{4}\frac{\tilde\alpha_{e(m)}^2}{\tilde\gamma_{e(m)}^2},
	\quad\qquad\ \quad  s^e_{L13}\simeq -\frac{\sqrt{3}}{3}|\epsilon_1^*|\,.
	\end{align}
	
	We can also obtain the mixing angles of the right-handed sector.
	Performing  the   unitary transformation $E_R\to U_{12}^T  E_R$
	and $\bar E_R\to \bar E_R U_{12}$
	for the case of  $\tilde\gamma_{e(m)}\gg\tilde\alpha_{e(m)}\gg\tilde\beta_{e(m)}$,
	where
		\begin{align}
	U_{12}=
	\begin{pmatrix}
	0 & 1 & 0\\  1& 0 & 0 \\ 0 & 0 &1
	\end{pmatrix}\,.
	\label{U12}
	\end{align}
Then, the charged lepton mass matrix in Eq.\eqref{charged-matrix-S}
turns to:
\begin{align} 
& M_{E}\,=\frac12 v_d
\begin{pmatrix}
0 & -3(\sqrt{3}-1) \tilde\beta_{e(m)} & -(3-\sqrt{3}) \tilde\beta_{e(m)}\\
0& 3(\sqrt{3}-1) \tilde\alpha_{e(m)} &-(3-\sqrt{3}) \tilde\alpha_{e(m)}\\
0 &0& 2(3-\sqrt{3})\tilde\gamma_{d(m)}
\end{pmatrix}_{RL}\,,
\end{align}
at $\tau=i$ .
The matrix 	$M_E M_E^\dagger $ is given at $\tau=i+\epsilon$ as: 
\begin{align} 
M_{E}M_{E}^\dagger \simeq 3(2-\sqrt{3})
\left [1+\frac23(3-\sqrt{3})\epsilon_1\right ]v_d^2
\begin{pmatrix}
2\tilde\beta_{e(m)}^2
&  -\tilde\alpha_{e(m)}\tilde\beta_{e(m)} 
&  -\tilde\beta_{e(m)}\tilde\gamma_{e(m)} \\
-\tilde\alpha_{e(m)}\tilde\beta_{e(m)}   &    2\tilde\alpha_{e(m)}^2 &
-\tilde\alpha_{e(m)}\tilde\gamma_{e(m)}\\
-\tilde\beta_{e(m)}\tilde\gamma_{e(m)} &   -\tilde\alpha_{e(m)}\tilde\gamma_{e(m)} & 
2\tilde\gamma_{e(m)}^2
\end{pmatrix}\,,
\label{charged-matrix-m}
\end{align} 
where $\tilde\gamma_{e(m)}\gg\tilde\alpha_{e(m)}\gg\tilde\beta_{e(m)}$.
	
The right-handed mixing angles are given as:
	\begin{align}
	U_{Rme}^\dagger M_E M_{E}^\dagger U_{Rme}={\rm diag}(m_e^2, m_\mu^2, m_\tau^2)\,,
	\qquad\quad 
	U_{Rme}\simeq 
	\begin{pmatrix}
	1 & s_{R12}^{e} & s_{R13}^{e}\\
	-s_{R12}^{e} & 1 & s_{R23}^{e} \\
	s_{R12}^e s_{R23}^e-s_{R13}^{e} & -s_{R23}^{e} & 1
	\end{pmatrix}
	\label{URde}\,,
	\end{align}
	where
	\begin{align}
	s^e_{R12}\simeq -\frac{\tilde\beta_{e(m)}}{\tilde\alpha_{e(m)}}\,,
	\qquad\quad   s^e_{R23}\simeq -\frac12\frac{\tilde\alpha_{e(m)}}{\tilde\gamma_{e(m)}}\,,\qquad\quad\ \  s^e_{R13}\simeq -\frac12\frac{\tilde\beta_{e(m)}}{\tilde\gamma_{e(m)}}\,.
	\end{align}



\end{document}